\newcommand{\beq}{\begin{equation}}
\newcommand{\eeq}{\end{equation}}
\newcommand{\bea}{\begin{eqnarray}}
\newcommand{\eea}{\end{eqnarray}}

\newcommand{\gsim}{\lower.7ex\hbox{$\;\stackrel{\textstyle>}{\sim}\;$}}
\newcommand{\lsim}{\lower.7ex\hbox{$\;\stackrel{\textstyle<}{\sim}\;$}}

\documentclass[aps,prev,onecolumn,preprintnumbers,floatfix,nofootinbib]{revtex4-1}
\usepackage{geometry,amsmath,amsfonts}
\usepackage{slashed}
\usepackage{graphicx}
\usepackage{epstopdf}
\usepackage{mathrsfs}
\usepackage{amssymb}
\usepackage{verbatim}
\usepackage{color}
\usepackage{multirow}
\usepackage{subfig}

\def\stacksymbols #1#2#3#4{\def\theguybelow{#2}
    \def\vp{\lower#3pt}
    \def\sp{\baselineskip0pt\lineskip#4pt}
    \mathrel{\mathpalette\intermediary#1}}

\def\intermediary#1#2{\vp\vbox{\sp
     \everycr={}\tabskip0pt
     \halign{$\mathsurround0pt#1\hfil##\hfil$\crcr#2\crcr
              \theguybelow\crcr}}}

\def\be{\begin{equation}}
\def\ee{\end{equation}}
\def\bea{\begin{eqnarray}}
\def\eea{\end{eqnarray}}

\def\sp{\;\;\;,\;\;\;}

\def\lsim{\raise0.3ex\hbox{$\;<$\kern-0.75em\raise-1.1ex\hbox{$\sim\;$}}}
\def\gsim{\raise0.3ex\hbox{$\;>$\kern-0.75em\raise-1.1ex\hbox{$\sim\;$}}}

\def\inbar{\,\vrule height1.5ex width.4pt depth0pt}

\def\IC{\relax\hbox{$\inbar\kern-.3em{\rm C}$}}
\def\IQ{\relax\hbox{$\inbar\kern-.3em{\rm Q}$}}
\def\IR{\relax{\rm I\kern-.18em R}}
 \font\cmss=cmss10 \font\cmsss=cmss10 at 7pt
\def\IZ{\relax\ifmmode\mathchoice
 {\hbox{\cmss Z\kern-.4em Z}}{\hbox{\cmss Z\kern-.4em Z}}
 {\lower.9pt\hbox{\cmsss Z\kern-.4em Z}}
 {\lower1.2pt\hbox{\cmsss Z\kern-.4em Z}}\else{\cmss Z\kern-.4em Z}\fi}

\def\comment#1{}
\def\to{\rightarrow}

\def\u1x{U(1)_X}
\newcommand{\nc}{\newcommand}
\nc{\LL}{L}
\nc{\vv}{\tilde{v}}
\nc{\ccdot}{\!\cdot\!}
\nc{\gsm}{G_{SM}}
\nc{\vfive}{\mathbf{5}\oplus\mathbf{\overline{5}}}
\nc{\vten}{\mathbf{10}\oplus\mathbf{\overline{10}}}
\nc{\zhol}{Z^{\rm hol}}
\nc{\xfb}{\,{\rm fb}}

\begin{document}

\title{2HDM portal for Singlet-Doublet Dark Matter}

\author{Giorgio Arcadi$^{a}$}
\email{arcadi@mpi-hd.mpg.de}

\vspace{0.1cm}
 \affiliation{
${}^a$ 
Max\--Planck\--Institut f\"ur Kernphysik (MPIK), Saupfercheckweg 1, 69117 Heidelberg, Germany
}

\begin{abstract} 
We present an extensive analysis of a model in which the (Majorana) Dark Matter candidate is a mixture between a SU(2) singlet and two SU(2) doublets. This kind of setup takes the name of singlet-doublet model. We will investigate in detail an extension of this model in which the Dark Matter sector interactions with a 2-doublet Higgs sector enforcing the complementarity between Dark Matter phenomenology and searches of extra Higgs bosons.
\end{abstract}

\maketitle

\section{Introduction}

\noindent
The WIMP paradigm is a compelling solution of the Dark Matter (DM) problem. It relates the achievement of the DM relic density, as measured with unprecedented precision by the PLANCK collaboration~\cite{Ade:2015xua}, to a specific range of values of the thermally averaged pair annihilation cross-section of the DM. An implication of this setup is that the DM should possess sizable interactions with the Standard Model (SM) particles, making possible a detection at present experimental facilities.

\noindent
From the model building perspective, a rather simple realization of the interactions needed by the WIMP paradigm consists into the existence of an electrically neutral mediator, coupled with pairs of DM states as well as pair of SM particles. The Higgs boson is a privileged candidate for this role~~\cite{Silveira:1985rk,
McDonald:1993ex,Burgess:2000yq,Kim:2006af,Andreas:2010dz,Kanemura:2010sh,Lebedev:2011iq,Mambrini:2011ik,
Djouadi:2011aa,LopezHonorez:2012kv,
Djouadi:2012zc,Cline:2013gha,Cornell:2016gho,Dick:2018lqx}. 

\noindent
A fermionic DM candidate, if it is a SM singlet, can couple, in pairs, with the Higgs boson only through $D>4$ operators. The so called singlet-doublet models~\cite{Cohen:2011ec,Cheung:2013dua,Yaguna:2015mva,Calibbi:2015nha}~\footnote{An extended version of these setups has been recently proposed to account for neutrino masses as well~\cite{Esch:2018ccs}.} overcome this problem by enlarging the specturm on BSM states by two to $SU(2)$ doublets, so that the DM is a mixture of their neutral components as well as of the singlet originally introduced. This has, however, the consequence that the DM can interact, in pairs, also with the $Z$ bosons, as well as, through the charged component of the extra doublets, with the $W$ bosons.

\noindent
As recently reviewed in~\cite{Arcadi:2017kky} (see also e.g.~\cite{Escudero:2016gzx,Balazs:2017ple}) DM interactions mediated exclusively by the Higgs and the $Z$ bosons are disfavored by DM Direct Detection (DD), expecially in the case that the DM is a dirac fermion~\cite{Arcadi:2014lta,Yaguna:2015mva,Angelescu:2016mhl}.

\noindent
In this work we will investigate in detail whether this problem can be encompassed by extending, with a second doublet, the Higgs sector of the theory. A similar investigation has been already presented in~\cite{Berlin:2015wwa}, but with focus only on the possibility of a light pseudoscalar boson. While including this scenario in our discussion, we will, however, investigate the parameter space of the theory from a more general perspective. We will pinpoint, furthermore, the complementarity with constraints from searches at collider and in low energy processes of extra Higgs bosons.

\noindent
The paper is organized as follows. We will first introduce, in section II, our model setup. Section III will be devoted to a brief review of the two-doublet extension of the Higgs sector and to the discussion of the theoretical and experimental limits which can impact the viable parameter space for DM. The most salient features of DM phenomenology  will be then discussed in section IV. We will finally present and discuss our findings in section V.

\section{The model}

\subsection{2HDM and coupling to the SM}

\noindent
We will adopt, for our study, a 2HDM model described by the the following potential:
\begin{align}
\label{eq:scalar_potential}
& V(H_1,H_2) = 
m_{11}^2 H_1^\dagger H_1+ m_{22}^2 H_2^\dagger H_2\nonumber\\
& - m_{12}^2 \left(H_1^\dagger H_2 + {\rm h.c.} \right)
+\frac{\lambda_1}{2} \left( H_1^\dagger H_1 \right)^2\nonumber\\
&+\frac{\lambda_2}{2} \left( H_2^\dagger H_2 \right)^2
\nonumber\\
& + \lambda_3 \left(H_1^\dagger H_1 \right)\left(H_2^\dagger H_2 \right)
+ \lambda_4 \left(H_1^\dagger H_2 \right)\left(H_2^\dagger H_1 \right)\nonumber\\
&+\frac{\lambda_5}{2}\left[ \left(H_1^\dagger H_2 \right)^2 + {\rm h.c.} \right],
\end{align}
where two doublets are defined by:
\begin{equation}
H_i=
\begin{pmatrix}
\phi_i^+ \\
(v_i+\rho_i +i \eta_i)/\sqrt{2}
\end{pmatrix}~,
\qquad
i=1,2,
\end{equation}
further assuming that all the couplings above are real. We assume, in addition, the existence of a discrete symmetry forbidding two additional couplings $\lambda_6$ and $\lambda_7$~\cite{Davidson:2005cw}. We also introduce, as usual, the $\beta$ angle defined as $v_2/v_1= \tan \beta$.

\noindent
Imposing CP-conservation, in the scalar sector, the spectrum of physical states is constituted by two CP even neutral states, $h$, identified with the 125 GeV Higgs, and $H$, the CP-odd Higgs $A$ and finally the charged Higgs $H^{\pm}$. The transition from the interaction basis $(H_1,H_2)^T$ to the mass basis $(h,H,A,H^{\pm})$ depends on two mixing angles, $\alpha$ and $\beta$.

\noindent
The quartic couplings of the scalar potential~(\ref{eq:scalar_potential}) can be expressed as function of the masses the physical states as:
\begin{align}
\label{eq:quartic_physical}
\lambda_1 &= \frac{1}{v^2} \left[ m_h^2 + \left( m_H^2 - M^2 \right) t_{\beta}^2 \right], \\
\lambda_2 &= \frac{1}{v^2} \left[ m_h^2 + \left( m_H^2 - M^2 \right) t_{\beta}^{-2} \right], \\
\lambda_3 &= \frac{1}{v^2} \left[ m_h^2 + 2 m_{H^{\pm}}^2 - \left( m_H^2 + M^2 \right) \right], \\
\lambda_4 &= \frac{1}{v^2} \left[ M^2 + m_A^2 - 2 m_{H^{\pm}} \right], \\
\lambda_5 &= \frac{1}{v^2} \left[ M^2 - m_A^2 \right], 
\end{align} 
where $M \equiv m_{12}/(s_{\beta} c_{\beta})$.

\noindent
SM fermions cannot couple freely with both Higgs doublets since, otherwise, FCNCs would be induced at tree level. Four specific configurations, labelled type I, type II, Lepton specific and flipped, (summarized in tab.~\ref{table:2hdm_type}) avoid this eventuality. In the physical basis for the scalar sector the interaction lagrangian between the Higgses and the SM fermions reads:
\begin{align}
-{\cal L}_{yuk}^{SM} &=\sum\limits_{f=u,d,l} \frac{m_f}{v} \left[\xi^f_h \overline{f}f h+\xi^f_H \overline{f}f H-i \xi^f_A \overline{f}\gamma_5 f A \right]
\notag
\\
&-\left[\frac{\sqrt{2}}{v} \overline{u} \left(m_u \xi^u_A P_L + m_d \xi^d_A P_R \right)d H^+ +\frac{\sqrt{2}}{v} m_l \xi_A^l \overline{\nu_L} l_R H^+  + \mathrm{h.c.} \right],
\end{align}
\noindent
where $v=\sqrt{v_1^2+v_2^2}=246\,\mbox{GeV}$. The coefficients $\xi_{h,H,A}^{u,d,l}$ are,  in general functions of $\alpha,\beta$, and depend on how the SM fermions are coupled with the $H_1,H_2$ doublets.

\begin{table}[h!]
\resizebox{1.0\textwidth}{!}{\begin{minipage}{\textwidth}
\begin{tabular}{|c|c|c|c|c|}
\hline
 &  Type I & Type II & Lepton-specific & Flipped \\
\hline\hline 
$\xi_h^u$ & $c_{\alpha}/ s_{\beta}$ & $c_{\alpha}/ s_{\beta} \rightarrow 1$ & $c_{\alpha}/ s_{\beta} \rightarrow 1$ & $c_{\alpha}/ s_{\beta}\rightarrow 1$
\\
\hline
$\xi_h^d$ & $c_{\alpha}/ s_{\beta} \rightarrow 1$ & $-s_{\alpha}/ c_{\beta} \rightarrow 1$ & $c_{\alpha}/ s_{\beta} \rightarrow 1$ & $-s_{\alpha}/ c_{\beta} \rightarrow 1$
\\
\hline
$\xi_h^l$ & $c_{\alpha}/ s_{\beta} \rightarrow 1$ & $-s_{\alpha}/ c_{\beta} \rightarrow 1$ & $-s_{\alpha}/ c_{\beta} \rightarrow 1$ & $c_{\alpha}/ s_{\beta} \rightarrow 1$ 
\\
\hline\hline
$\xi_H^u$ & $s_{\alpha}/ s_{\beta} \rightarrow -t_{\beta}^{-1}$ & $s_{\alpha}/ s_{\beta} \rightarrow -t_{\beta}^{-1}$ & $s_{\alpha}/ s_{\beta} \rightarrow -t_{\beta}^{-1}$ & $s_{\alpha}/ s_{\beta} \rightarrow -t_{\beta}^{-1}$
\\
\hline
$\xi_H^d$ & $s_{\alpha}/ s_{\beta} \rightarrow -t_{\beta}^{-1}$ & $c_{\alpha}/ c_{\beta} \rightarrow t_{\beta}$ & $s_{\alpha}/ s_{\beta} \rightarrow -t_{\beta}^{-1}$ & $c_{\alpha}/ c_{\beta} \rightarrow t_{\beta}$
\\
\hline
$\xi_H^l$ & $s_{\alpha}/ s_{\beta} \rightarrow -t_{\beta}^{-1}$ & $c_{\alpha}/ c_{\beta} \rightarrow t_{\beta}$ & $c_{\alpha}/ c_{\beta} \rightarrow t_{\beta}$ & $s_{\alpha}/ s_{\beta} \rightarrow -t_{\beta}^{-1}$
\\
\hline\hline
$\xi_A^u$ & $t_{\beta}^{-1}$ & $t_{\beta}^{-1}$ & $t_{\beta}^{-1}$ & $t_{\beta}^{-1}$
\\
\hline
$\xi_A^d$ & $-t_{\beta}^{-1}$ & $t_{\beta}$ & $-t_{\beta}^{-1}$ & $t_{\beta}$
\\
\hline
$\xi_A^l$ & $-t_{\beta}^{-1}$ & $t_{\beta}$ & $t_{\beta}$ & $-t_{\beta}^{-1}$
\\
\hline
\end{tabular}
\caption{\small Couplings of the Higgses to the SM fermions as a function of the \\ angles $\alpha$ and $\beta$ and in the alignment limit where $(\beta-\alpha) \rightarrow \pi/2$.}
\label{table:2hdm_type}
\end{minipage}}
\end{table}

\noindent
Constraints from 125 GeV Higgs signal strengh limit the values of $\alpha$ and $\beta$. These bounds will be discussed in more detail below. We just mention that one can automatically comply with them by going to the so-called ``alignment'' limit, i.e. $\beta -\alpha =\frac{\pi}{2}$, which makes automatically the couplings of the $h$ state SM like, i.e. $\xi_h^{u,d,l}=1$, and the other $\xi$ parameters only dependent on $\tan\beta$. A further implication of the alignment limit is that the coupling of the CP-even state $H$ with the $W$ and $Z$ bosons is null (the couplings of the $A$ boson with a $ZZ$ and $WW$ is null as long CP is conserved in the Higgs sector). In our study we will keep as free as possible the parameters of the Higgs sector. We will then do not strictly impose the alignment limit but rather keep $\alpha$ and $\beta$ as free parameters and impose on them the relevant constraints.

\subsection{Coupling to the DM}

\noindent
In the scenario under investigation, the DM arises from the mixture of a SM singlet $N^{'}$ and the neutral components of two (Weyl) $SU(2)$ doublets $L_L$ and $L_R$ defined as:

\begin{equation}
L_L=\left(
\begin{array}{c}
N_L \\
E_L
\end{array}
\right),\,\,\,\,\,
L_R=\left(
\begin{array}{c}
-E_R \\
N_R
\end{array}
\right),
\end{equation}

\noindent
The new fermions are coupled with two Higgs doublets $H_1$ and $H_2$ according the following lagrangian:
\begin{equation}
\mathcal{L}=-\frac{1}{2}M_N N^{'\,2}-M_L L_L L_R -y^L_i L_L H_i N^{'}-y_i^R \bar N^{'} \tilde{H}_i^{\dagger} L_R  +\mbox{h.c.},\,\,\,i=1,2
\end{equation}

\noindent
Notice that, given their quantum numbers, the new fermions could be coupled, through the Higgs, also with SM leptons and, hence, mix with them after EW symmetry breaking. This would imply, in particular, that the DM is not stable. To avoid this possibility we assume the existence of a discrete symmetry under the new and the SM fermions are, respectively, odd and even.

\noindent
Similarly to the case of SM fermions, coupling the new states freely with both Higgs doublets might lead to FCNC (this time at loop level). This problem can be overcome by considering similar coupling configurations as the ones of SM fermions. 

\noindent
Along this work we will focus on two configurations~\footnote{These configurations can be enforced by charging (at least some of) the new fermions and the Higgs doublets under suitable $Z_2$ symmetry (see e.g.~\cite{Angelescu:2016mhl} for a discussion).}:$y_1^L=y_1$, $y_2^L=y_1^R=0$, $y_2^R=y_2$ and  $y_1^L=y_1$, $y_2^L=y_2^R=0$, $y_1^R=y_2$. For the first configuration we will further assume that the SM fermions are coupled only with the $H_1$ doublet and globally label as ``Type-I'' the model defined in this way. According an analogous philosophy we assume that the second configuration for the new fermions is accompanied by coupling of the SM fermions with $H_1,H_2$ as in the type-II 2HDM and define as ``type-II' this scenario. 

\noindent
After EW symmetry breaking mixing between $N'$ and the neutral componenents of $L_L$ and $L_R$ occurs, so that the physical spectrum of the new fermions is represented by three Majorana fermions $\psi_{i=1,3}$ defined by:
\begin{equation}
\psi_i=N^{'} U_{i1}+N_L U_{i2}+N_R U_{i3},
\end{equation}
and one charged dirac fermion $\psi_{\pm}$ with mass $m_{\psi_{\pm}} \approx M_L$. The matrix $U$ diagonalizes a mass matrix of the form:
\begin{equation}
M=\left(
\begin{array}{ccc}
M_N & \frac{y_1 v_1}{\sqrt{2}} & \frac{y_2 v_2}{\sqrt{2}} \\
\frac{y_1 v_1}{\sqrt{2}} & 0 & M_L \\
\frac{y_2 v_2}{\sqrt{2}} & M_L & 0
\end{array}
\right),
\end{equation}
where, for definiteness we have considered the type-II model. It can be easily argued that the mass matrix above resembles the bino-higgsino (or singlino-higgino) mass matrix in MSSM (NMSSM) scenario once one identifies $M_N$ and $M_L$ with, respectively, the Majorna mass $M_1$ of the Bino and the supersymmetric $\mu$ parameter, and performs the substitution $y \rightarrow g'/\sqrt{2}$ with $g'$ being the hypercharge gauge coupling. 

\noindent
Similarly to~\cite{Cheung:2013dua,Calibbi:2015nha,Berlin:2015wwa}, we will adopt, in spite of $y_1,y_2$, the free parameters $y,\theta$ defined by:
\begin{equation}
y_1=y \cos\theta\,\,\,\,y_2=y \sin\theta
\end{equation}

\noindent
In the mass basis the relevant interaction lagrangian for DM phenomenology reads:
\begin{align}
& \mathcal{L}=\overline{\psi^-} \gamma^\mu \left(g^V_{W\chi_i}-g^A_{W\chi_i}\gamma_5\right)\psi_i W_\mu^{-}+\mbox{h.c.}
+\frac{1}{2}\sum_{i,j=1}^3 \overline{\psi_i} \left(g_{Z \psi_i \psi_j}^V-g_{Z \psi_i \psi_j}^A \gamma_5\right) \psi_j \nonumber\\
& +\frac{1}{2}\sum_{i,j=1}^{3}\overline{\psi_i}\left(y_{h \psi_i \psi_j}h+y_{H \psi_i \psi_j}H+y_{A \psi_i \psi_j}\gamma_5 A\right)\psi_j+\mbox{h.c}\nonumber\\
& +\overline{\psi^-} \left(g^S_{H^{\pm}\psi_i}-g^P_{H^{\pm}\psi_i}\gamma_5\right)\psi_i W_\mu^{-}+\mbox{h.c.}
\end{align}
with:
\begin{align}
& g^V_{W\chi_i}=\frac{g}{2\sqrt{2}}\left(U_{i3}-U_{i2}^{*}\right)\nonumber\\
& g^A_{W\chi_i}=-\frac{g}{2\sqrt{2}}\left(U_{i3}+U_{i2}^{*}\right) \nonumber\\
& g^V_{Z\psi_i \psi_j}=\frac{g}{4 \cos\theta_W}\left[\left(U_{i3}U_{j3}^{*}-U_{i2}U_{j2}^{*}\right)-\left(U^{*}_{i3}U_{j3}-U_{i2}^{*}U_{j2}\right)\right]\nonumber\\
& g^A_{Z\psi_i \psi_j}=\frac{g}{4 \cos\theta_W}\left[\left(U_{i3}U_{j3}^{*}-U_{i2}U_{j2}^{*}\right)+\left(U^{*}_{i3}U_{j3}-U_{i2}^{*}U_{j2}\right)\right]\nonumber\\
& g_{h\psi_i \psi_j}=-\frac{1}{2\sqrt{2}}\left[U_{i1}\left(y_1 N_1^h U_{i2}+y_2 N_2^h U_{i3}\right)+(i \leftrightarrow j)\right] \nonumber\\
& g_{H\psi_i \psi_j}=-\frac{1}{2\sqrt{2}}\left[U_{i1}\left(y_1 N_1^H U_{i2}+y_2 N_2^H U_{i3}\right)+(i \leftrightarrow j)\right] \nonumber\\
& g_{A\psi_i \psi_j}=-\frac{i}{2\sqrt{2}}\left[U_{i1}\left(y_1 N_1^A U_{i2}+y_2 N_2^A U_{i3}\right)+(i \leftrightarrow j)\right] \nonumber\\
& g^S_{H^{\pm}\psi_i}=\frac{1}{2}U_{i1}\left(y_1 N_1^{H^{\pm}}+y_2 N_2^ {H^{\pm}}\right)\nonumber\\
& g^P_{H^{\pm}\psi_i}=\frac{1}{2}U_{i1}\left(y_1 N_1^{H^{\pm}}-y_2 N_2^ {H^{\pm}}\right)\nonumber\\
\end{align}

where:
\begin{align}
& N_1^h=N_2^h=-\sin\alpha,\,\,\,N_1^H=N_2^H=\cos\alpha,\,\,\,N_1^A=N_2^A=-\sin\beta,\,\,\,N_1^{H^{\pm}}=N_2^{H^{\pm}}=-\sin\beta\,\,\,\mbox{(Type-I)}\nonumber\\
& N_1^h=-\sin\alpha,\,\,\,\,N_2^h=\cos\alpha,\,\,\,\,N_1^H=\cos\alpha,\,\,\,\,N_2^H=\sin\alpha\nonumber\\
& N_1^A=-\sin\beta\,\,\,\,\,N_2^A=\cos\beta,\,\,\,\,N_1^{H^{\pm}}=-\sin\beta,\,\,\,\,N_2^{H^{\pm}}=\cos\beta\,\,\,\,\mbox{(Type-II)}
\end{align}

\noindent
Notice, in particular, that, as expected from its Majorana nature, the vectorial coupling of the DM with the $Z$ boson is null. 

\section{Constraints on the Higgs sector}

\subsection{Bounds on the potential}

\noindent
The quartic couplings $\lambda_{i=1,5}$ should comply with a series of constraints coming from the unitarity and boundness from below of the scalar potential as well as perturbativity (see for example.~\cite{Kanemura:2004mg,Becirevic:2015fmu} for more detailed discussions). These bounds, through eq.~\ref{eq:quartic_physical}, can be translated into bounds on the masses of the new Higgs bosons as function of the angles $\alpha$ and $\beta$. 

\noindent
For completeness we list below the main constraints:
\begin{itemize}
\item Scalar potential bounded from below:

\begin{equation}
\label{eq:up1}
\lambda_{1,2} > 0, \; \lambda_3 > -\sqrt{\lambda_1\lambda_2}, \; {\rm and} \; \lambda_3 + \lambda_4 - \left|\lambda_5\right| > -\sqrt{\lambda_1\lambda_2};
\end{equation}
\item tree level s-wave unitarity:

\beq
\label{eq:up2}
\left| a_{\pm} \right|, \left| b_{\pm} \right|, \left| c_{\pm} \right|, \left| f_{\pm} \right|, \left| e_{1,2} \right|, \left| f_1 \right|, \left| p_1 \right| < 8\pi,
\eeq
where:
\begin{align}
a_{\pm} &= \frac{3}{2}(\lambda_1 + \lambda_2) \pm \sqrt{\frac{9}{4}(\lambda_1-\lambda_2)^2 + (2\lambda_3 + \lambda_4)^2}, \notag \\
b_{\pm} &= \frac{1}{2}(\lambda_1 + \lambda_2) \pm \sqrt{(\lambda_1-\lambda_2)^2 + 4\lambda_4^2}, \notag \\
c_{\pm} &= \frac{1}{2}(\lambda_1 + \lambda_2) \pm \sqrt{(\lambda_1-\lambda_2)^2 + 4\lambda_5^2}, \notag \\
e_1 &= \lambda_3 + 2\lambda_4 - 3\lambda_5, \quad e_2 = \lambda_3 - \lambda_5, \notag \\
f_+ &= \lambda_3 + 2\lambda_4 + 3\lambda_5, \!\quad f_- = \lambda_3 + \lambda_5, \notag \\
f_1 &= \lambda_3 + \lambda_4, \quad p_1 = \lambda_3 - \lambda_4;
\end{align}
\item vacuum stability:
\begin{equation}
\label{eq:vacuum_2HDM}
m_{12}^2 \left(m_{11}^2-m_{22}^2 \sqrt{\lambda_1/\lambda_2}\right)\left(\tan\beta-\sqrt[4]{\lambda_1/\lambda_2}\right)>0
\end{equation}
\noindent
where the mass paramaters $m_{11},m_{22},m_{12}$ should satisfy:
\begin{align}
& m_{11}^2+\frac{\lambda_1 v^2 \cos^2\beta}{2}+\frac{\lambda_3 v^2 \sin^2\beta}{2} =\tan\beta \left[m_{12}^2-(\lambda_4+\lambda_5)\frac{v^2 \sin 2\beta}{4}\right]\nonumber\\
& m_{22}^2+\frac{\lambda_2 v^2 \sin^2\beta}{2}+\frac{\lambda_3 v^2 \cos^2\beta}{2}=\frac{1}{\tan\beta} \left[m_{12}^2-(\lambda_4+\lambda_5)\frac{v^2 \sin2\beta}{4}\right].
\end{align}
\end{itemize}

\subsection{EWPT}

\noindent
The presence of extra Higgs bosons affects the values of the Electroweak precision observables, possibly making them to deviate from the SM expectations. While eventual tensions can be automatically alleviated imposing specific relations for the masses of the new states; for example deviations of the $T$ parameter can be avoided imposing a custodial symmetry~\cite{Davidson:2010xv,Baak:2011ze}, for our DM analysis, as already pointed, we will try keep the parameters of the Higgs sector as free as possible. To identify the viable paramter space we have then computed the $S,T,U$ parameters (see e.g.~\cite{Maksymyk:1993zm,Barbieri:2006bg,Branco:2011iw} for the corrisponding expressions)  and determining the excluded model configurations through the following $\chi^2$~\cite{Bertuzzo:2016fmv,Arnan:2017lxi,Becirevic:2017chd}:
\begin{equation}
\chi^2=\sum_{i,j}(X_i-X_i^{\rm SM}){\left(\sigma_i V_{ij} \sigma_j\right)}^{-1}(X_j-X_j^{\rm SM})
\end{equation}
with $X=(S,T,U)$ while $X^{\rm SM}$, $\sigma$ and $V$ represent, respectively, the SM expectation of the EWP parameters, the corresponding Standard deviation and covariant matrix. Their updated values have been provided here~\cite{Baak:2014ora} and are also reported here for convenience:
\begin{align}
& X^{\rm SM}=(0.05,0.09,0.01),\,\,\,\,\,\sigma=(0.11,0.13,0.11)\nonumber\\
&V=\left(
\begin{array}{ccc}
1 & 0.9 & -0.59 \\
0.9 & 1 & -0.83 \\
-0.59 & -0.83 & 1
\end{array}
\right)
\end{align}
\noindent
We have imposed to each model point to not induce a deviation, for the EWP parameters, beyond $3\sigma$ from the best fit values.

\subsection{Collider searches of the new Higgs bosons}

\noindent
$H$ and $A$ bosons can be resonantly produced at colliders through gluon fusion~\footnote{In some configurations also $\bar b b$ fusion can play a relevant role~\cite{Angelescu:2016mhl}. These configurations are not object of present study.} and are, hence, object of searches in broad variety of final states ranging from $\bar \tau \tau$~\cite{Aaboud:2017sjh,CMS-PAS-HIG-16-037} $\bar b b$~\cite{Khachatryan:2015tra,CMS-PAS-HIG-16-025,Aaboud:2017xsd,Sirunyan:2017uvf}, $\bar t t$~\cite{Aaboud:2017hnm}, diboson~\cite{CMS-PAS-HIG-16-033,CMS-PAS-HIG-16-034,Aaboud:2016okv,Aaboud:2017itg,CMS-PAS-HIG-16-023,Aaboud:2017fgj,Aaboud:2017gsl,Aaboud:2017yyg}, $hh$ (only for $H$)~\cite{ATLAS-CONF-2016-071,CMS-PAS-HIG-16-002,CMS-PAS-HIG-16-032,ATLAS-CONF-2016-004,Sirunyan:2017djm,Sirunyan:2017guj}, $Zh$ (only for $A$)~\cite{ATLAS-CONF-2016-015,Aad:2015wra,Khachatryan:2015tha},  $A \rightarrow ZH$ and $H \rightarrow ZA$~\cite{Khachatryan:2015lba,Khachatryan:2016are}.

\begin{figure}
\begin{center}
\subfloat{\includegraphics[width=0.5\linewidth]{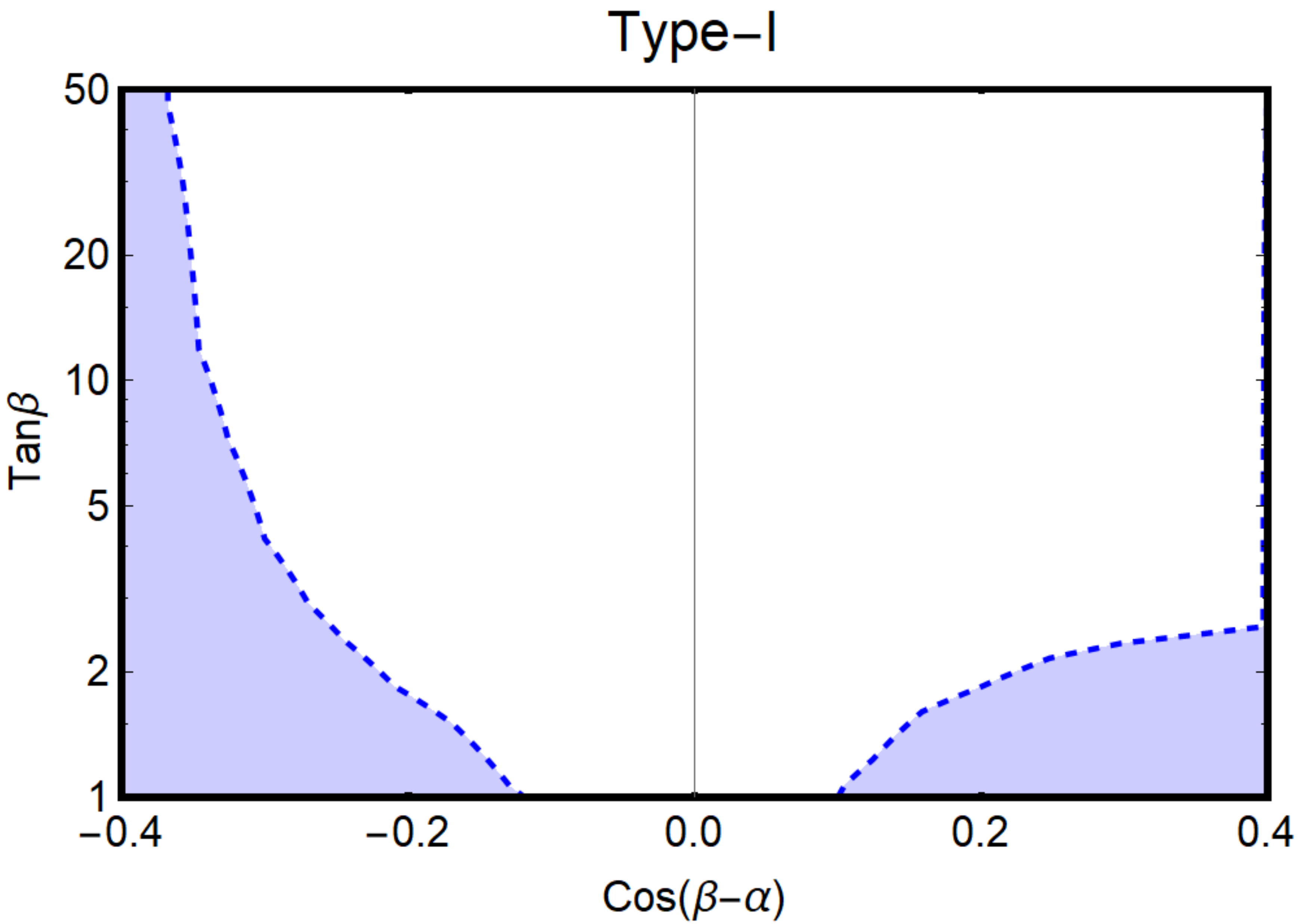}}
\subfloat{\includegraphics[width=0.5\linewidth]{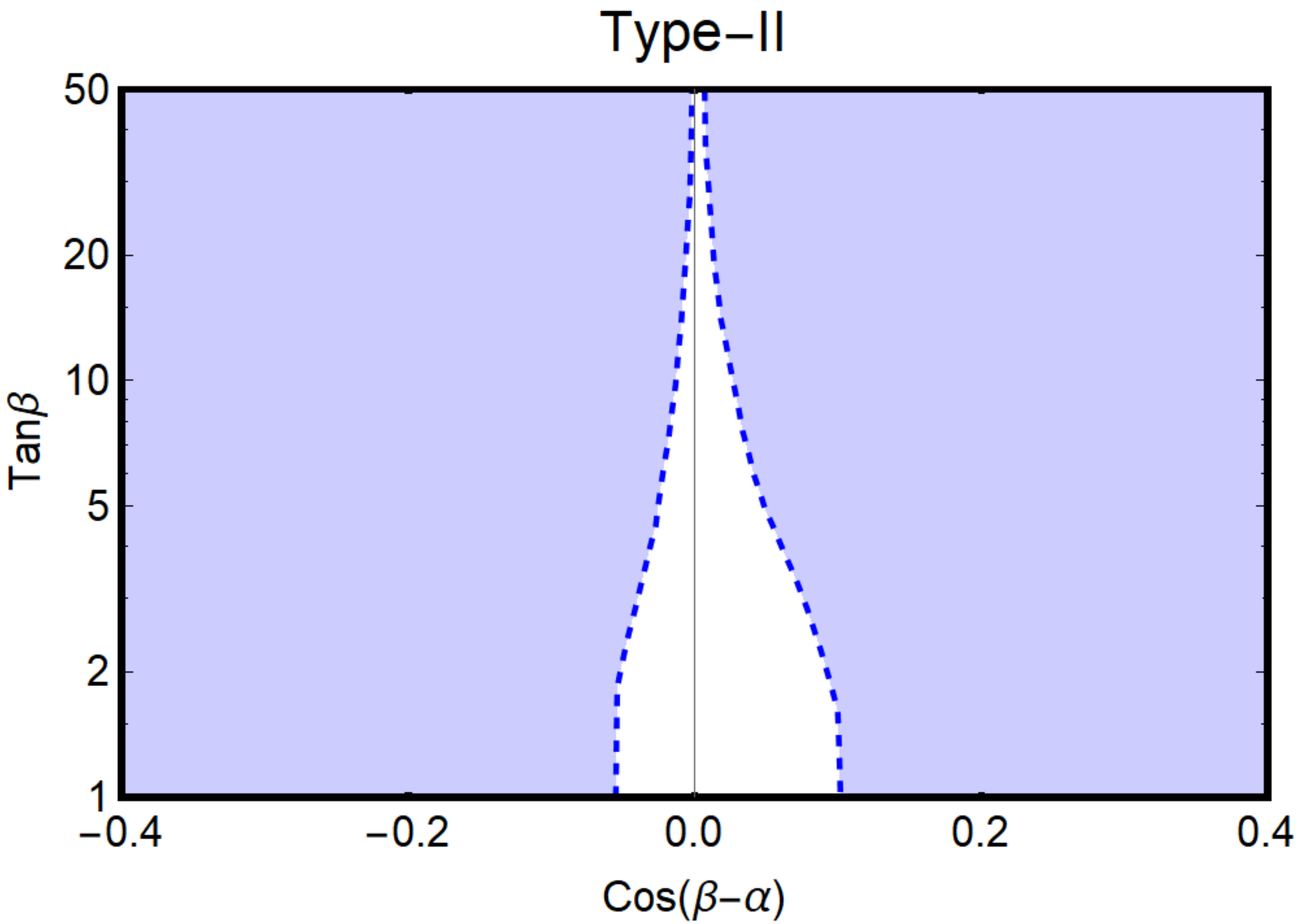}}
\end{center}
\caption{\footnotesize{Limits on deviations from the alignment limit in the for type-I (left panel) and type-II (right panel) 2HDM.}}
\label{fig:limital}
\end{figure}

\noindent
Among these, the most effective bounds come from searches of $\bar \tau \tau$ pairs, which exclude, for the type-II model, moderate-high values of $\tan\beta$, above 10. 
Searches of diboson final states and of the process $A \rightarrow Zh(H)$ (as well as the process with inverse mass ordering) can, in addition, constrain deviations from the alignment limit. The strongest limits for the latter come, however, from the Higgs signal strenght~\cite{Aad:2015gba,Khachatryan:2014jba,Bauer:2017fsw} as evidenced in fig.~\ref{fig:limital}.

\noindent
Similarly to~\cite{Berlin:2015wwa}, we will include in our analysis also the case in which $m_A < m_h$. In such a case one should take into account possible limits on the process $h \rightarrow AA$. This can be generically constrained through the Higgs signal strenght (i.e. one generically imposes that the branching fraction of this decay does not exceed the allowed value of the invisible branching fraction of the Higgs. This bound is effective in the whole range of masses for which the decay $h \rightarrow AA$ is kinematically allowed) as well as through dedicated searches~\cite{Khachatryan:2017mnf} (limits are effective only for some range of masses). More contrived are instead the prospect for signals not related to Higgs decays, see anyway~\cite{Bauer:2017ota,Goncalves:2016iyg}

\noindent
Concerning the charged Higgs boson we have first of all a limit $m_{H^{\pm}}\gtrsim 80\,\mbox{GeV}$ from LEP~\cite{Abbiendi:2013hk}. Moving to LHC constraints, these come, for $m_{H^{\pm}} < m_t$, from searches of top decays $t \rightarrow H^{\pm} b$ with $H^{\pm}$~\cite{Aad:2013hla,Aad:2014kga,Khachatryan:2015qxa,Khachatryan:2015uua} decaying into $\tau \nu_{\tau}$ or $cs$. The corresponding limits have been reformulated in~\cite{Arbey:2017gmh} for different realizations of the 2HDM. Values of the masses of the charged Higgs for which the decay is kinematically allowed are excluded for $\tan\beta \lesssim 10$ in the type-I 2HDM and irrespective of the value of $\tan\beta$ for the type-II scenario. For $m_{H^{\pm}}>m_t$ searches rely on direct production of the charged Higgs in association with a top and a bottom quark, followed by the decay of the former in $\tau\nu$~\cite{Khachatryan:2015qxa,Aaboud:2016dig,ATLAS-CONF-2016-089,CMS-PAS-HIG-16-031} or $tb$~\cite{ATLAS-CONF-2016-089}. Associated limits are not competitive with the others discussed in here and will be then neglected. The charged Higgs feels, indirectly, also limits from searches of the neutral Higgs bosons since the conditions on the quartic couplings~\ref{eq:up1}-\ref{eq:up2} impose relations between the masses of the new Higgs bosons (see e.g~\cite{Arbey:2017gmh}).

\subsection{Limits from flavour}

\noindent
While it is possible to avoid that the couplings of the extra Higgs bosons with SM fermions induce FCNC at the tree level, they can impact flavor violating transitions at the loop level. The strongest limits come from processes associated to $b \rightarrow s$ transitions. Their rates are mostly sensitive to $m_{H^{\pm}}$ and $\tan\beta$. Experimental limits are formulated in terms of these parameters. The most stringent come from the $B\rightarrow X_s \gamma$ processes~\cite{Amhis:2016xyh} and are particularly severe in the case of type-II model, excluding $m_{H^{\pm}}< 570\,\mbox{GeV}$ irrespective of $\tan\beta$~\cite{Misiak:2017bgg}. Much better is, instead, the situation of the type-I model where we have an approximate lower bound $\tan\beta \gtrsim 2$. A similar exclusion, for both type-I and type-II models is also provided by the processes $B_s \rightarrow \mu^+ \mu^-$ and $B \rightarrow K \mu^+ \mu^-$~\cite{Arnan:2017lxi}.

\subsection{Scanning the parameter space}

\noindent
In order to determine the allowed ranges of parameters of the Higgs sector, which can be interfaced with the DM sector of the theory, we have performed a scan over the following ranges:
\begin{align}
& \tan\beta \in [1,50]\,\,\,\,\alpha \in \left[-\frac{\pi}{2},\frac{\pi}{2}\right] \nonumber\\
& m_H=M \in [m_h,1\,\mbox{TeV}], \,\,\,\,m_{H^{\pm}}\in [m_W,1\,\mbox{TeV}]\,\,\,\,m_A \in [20\,\mbox{GeV},1\,\mbox{TeV}],
\end{align} 
imposing the constraints~\ref{eq:up1}-\ref{eq:vacuum_2HDM}, from the Higgs signal strenght (see fig.~\ref{fig:limital}), EWPT, as well the ones from searches of the Higgs bosons and from flavour physics. As already said we will include in our analysis a very light pseudoscalar $A$. For this reason we have considered a minimal value of 20 GeV for its mass in the scan.

\begin{figure}
\begin{center}
\subfloat{\includegraphics[width=0.5\linewidth]{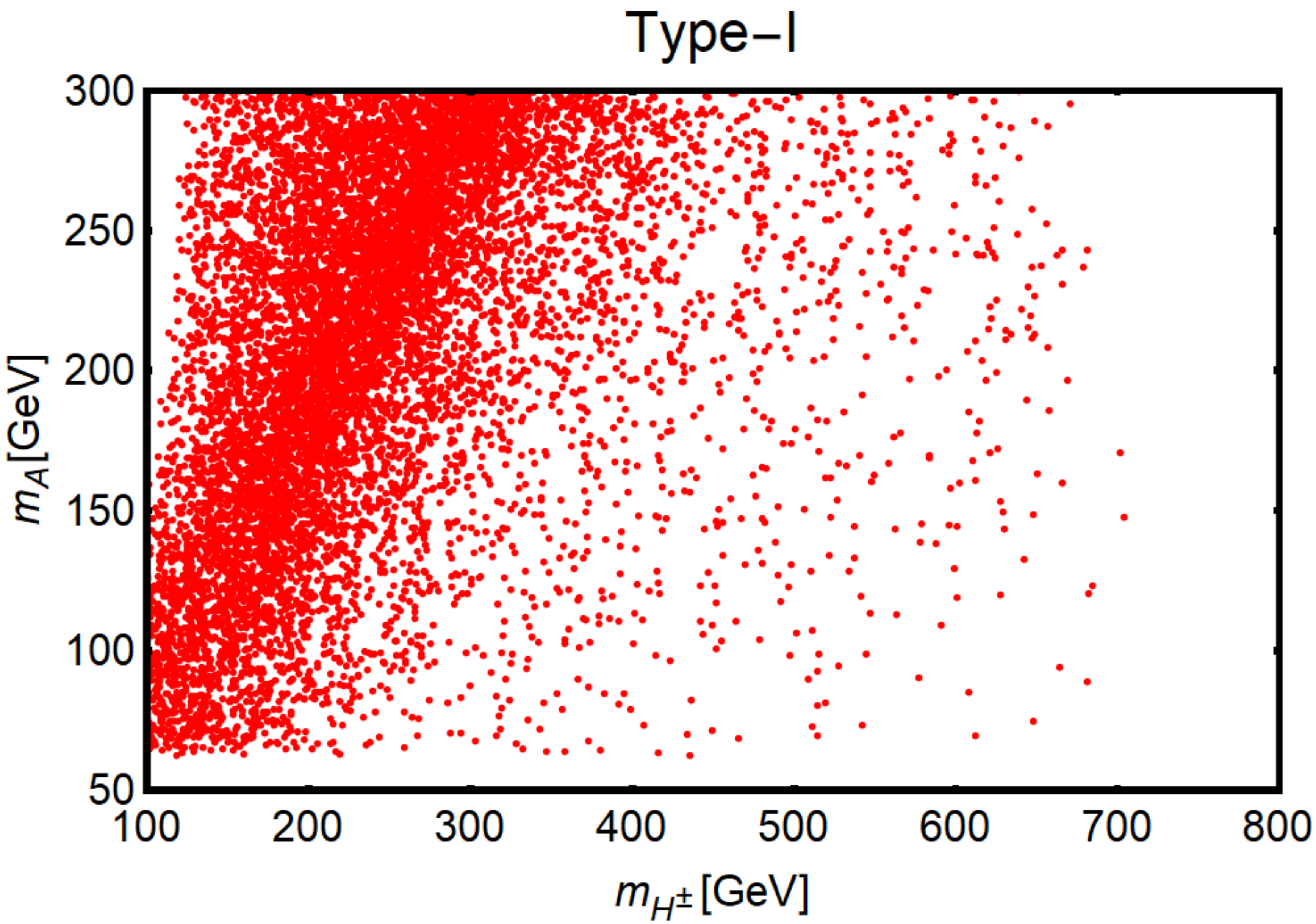}}
\subfloat{\includegraphics[width=0.5\linewidth]{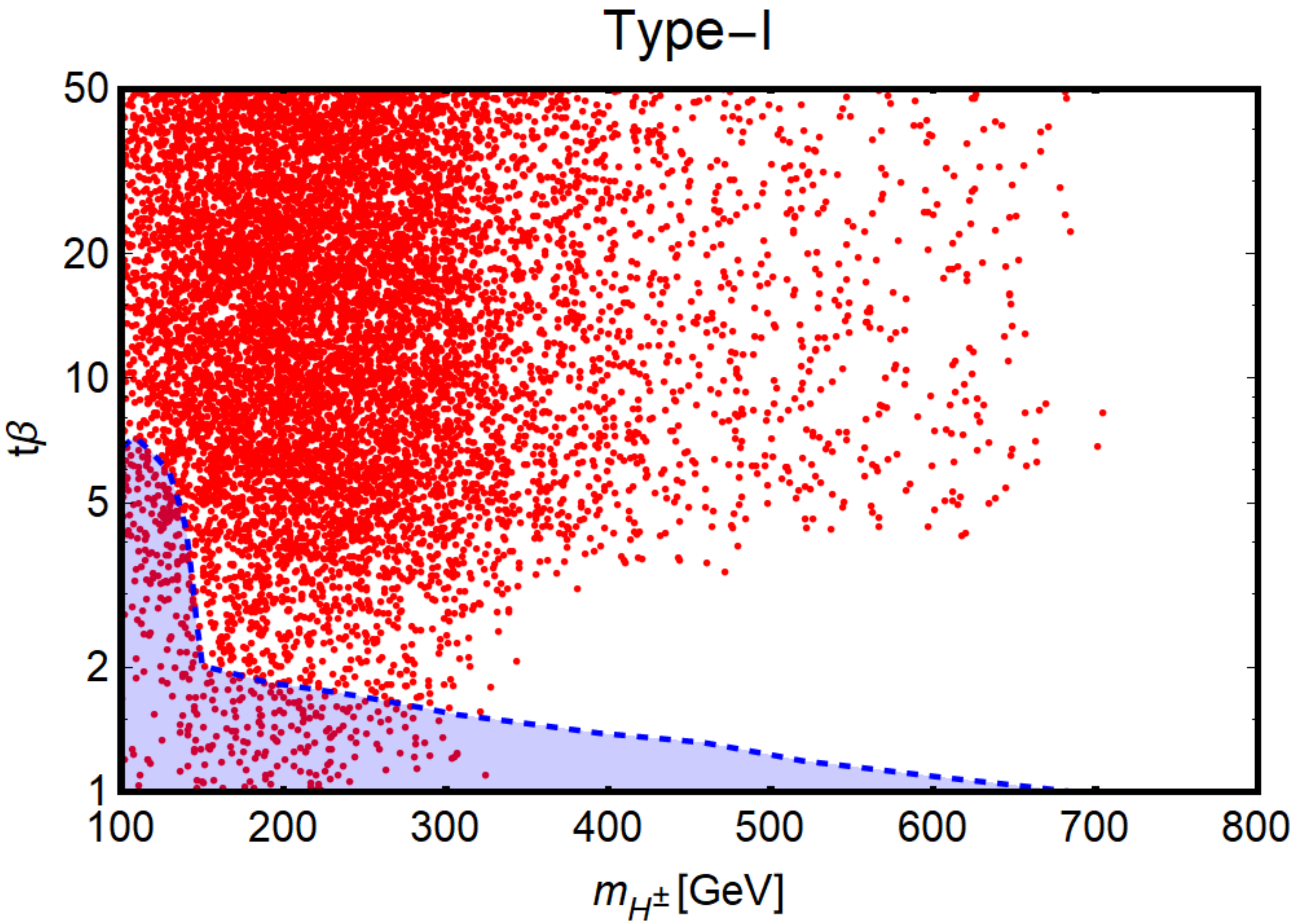}}
\end{center}
\caption{\footnotesize{Model points for the type-I model in the bidimensional plane $(m_{H^{\pm}},m_A)$ (left panel) and $(m_{H^{\pm}},t\beta)$ (right panel). The blue region in the right panel is excluded by bounds form low energy processes.}}
\label{scantI}
\end{figure}

\begin{figure}
\begin{center}
\subfloat{\includegraphics[width=0.5\linewidth]{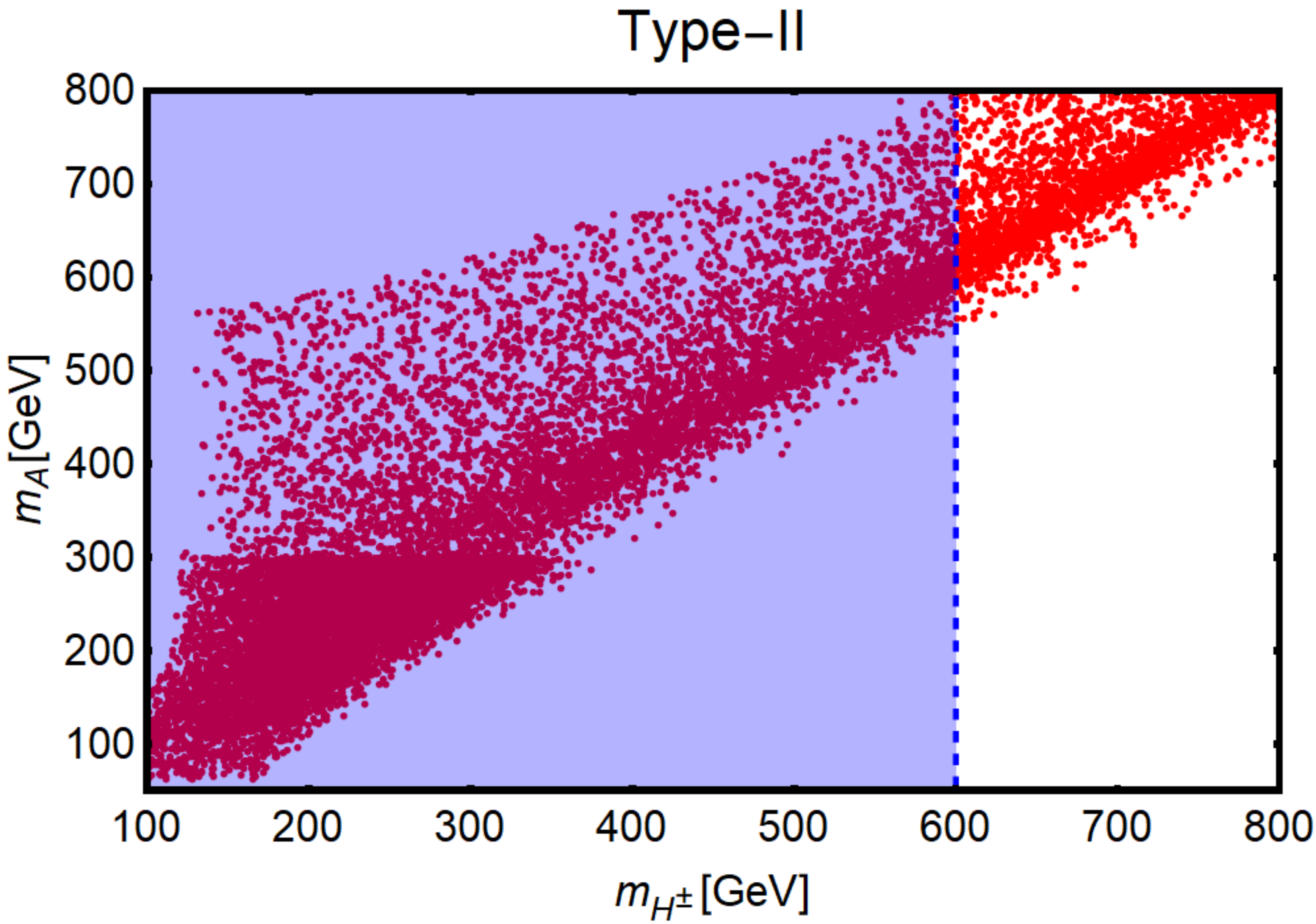}}
\end{center}
\caption{\footnotesize{Model points of the type-II 2HDM in the bidimensional plane $(m_{H^{\pm}},m_A)$. The blue region is excluded by flavor constraints (mostly $b \rightarrow s$ transitions.)}}
\label{scantII}
\end{figure}

\noindent
The most salient results for the type-I and type-II models are shown, respectively, in fig.~\ref{scantI} and fig~\ref{scantII}. 

\noindent
In the case of type-I model we have represented our results in the bidiminesional planes $(m_{H^{\pm}},m_A)$ and $(m_{H^{\pm}},\tan\beta)$. While the scan extented over larger ranges, we have highlighted, in the presentation of the results, the low mass region for $m_A$. The plots show, in particular, that it is possible, compatibly with the different constraints, to achieve a sizable hierarchy between the mass of the CP-odd Higgs $A$ and the one of the charged Higgs. The second panel shows, instead, the excluded region, mostly by flavor constraints.

\noindent
In the case of the type-II model the bound from $b \rightarrow s$ transitions is substantially independent from $\tan\beta$. We have then just presented the results in the bidimensional plane $(m_A,m_{H^{\pm}})$. As evidenced the second panel of fig.~\ref{fig:limital}, only very tiny deviation from the alignment limit are allowed. This implies, in turn, that the masses of the $H,A,H^{\pm}$ should lie relatively close each other. The strong exclusion bound on the mass $m_{H^{\pm}}$ from $Br(B \rightarrow X_s \gamma)$ hence corresponds to a lower bound of $m_A$ of the order of 500 GeV.

\section{DM constraints}

\subsection{Relic density}

\noindent
According the WIMP paradigm the DM has sizable enough interactions with the SM particles to be in thermal equilibrium in the Early stages of the history of the Universe. At a later stage the interaction rate of the DM fell below the Hubble expansion rate causing the freeze-out of the DM at temperatures of the order of $\frac{1}{20}-\frac{1}{30}$ the DM mass. Assuming standard cosmological history the DM relic density, $\Omega h^2$ is determined by a single particle physics input, i.e. the DM thermally averaged pair annihilation cross-section. The relation between these two quantities is given by~\cite{Gondolo:1990dk}:
\begin{equation}
\Omega_{\psi_1}h^2 \approx 8.76\times 10^{-11}\,{\mbox{GeV}}^{-2} \left[\int_{T_0}^{T_f} g_{*}^{1/2} \langle \sigma_{eff} v \rangle \frac{dT}{m_{\psi_1}}\right]
\end{equation}
where $T_0$ and $T_f$ represent, respectively, the present time and freeze-out temperature with the latter being typically $T_f \sim \frac{m_{\psi_1}}{20}-\frac{m_{\psi_1}}{30}$ while $g_{*}^{1/2}$ is a function of the relavistic degrees of freedom at the temperature $T$~\cite{Gondolo:1990dk}. $\langle \sigma_{\rm eff} v \rangle$ is the effective annihilation cross-section~\cite{Edsjo:1997bg}:
\begin{equation}
\label{eq:sigmacoann}
\langle \sigma_{\rm eff} v \rangle=\sum_{i,j \in \psi_{1,2,3},\psi^{\pm}} \langle \sigma_{ij} v_{ij} \rangle \frac{n_i}{n_{i,eq}}\frac{n_j}{n_{j,eq}}
\end{equation}
including coannihilation effects from the additional neutral and charged states belonging to the Dark Matter sector. Coannihilation effects are expected to be important in the case $M_L \lesssim M_N$, corresponding to a DM with sizable or even dominant doublet component, implying that at least the charged fermion $\psi^{\pm}$ is very close in mass to it.

\noindent
We remind that the precise experimental determination $\Omega h^2 \approx 0.12$~\cite{Ade:2015xua} is matched by a value of $\langle \sigma_{\rm eff} v \rangle$ of the order of $10^{-26}\,{\mbox{cm}^3}{\mbox{s}}^{-1}$. 

\noindent
In the model considered in this work a huge variety of processes contribute to $\langle \sigma v_{\rm eff} \rangle$. For what regards DM pair annihilations, the most commonly considered are the annihilation into SM fermion pairs, $WW$, $ZZ$ and $Zh$, originated by s-channel exchange of the $h,H,A$ bosons as well as, in the case of annihilation into gauge boson pairs, t-channel exchange of the DM and the other new fermions. In this work we will put particular attention also to the case in which some of new Higgs bosons, in particular the pseduscalar $A$, is light. This allows for additional annihilation channels in higgs bosons pairs, namely $HH$, $hA$ $HA$, $AA$, $H^+ H^-$, as well as gauge-higgs bosons final states as, $W^{\pm}H^{\mp}$, $ZH$ and $ZA$. As already mentioned, this broad collection of processes is further enriched by coannihilations, i.e. annihilation processes with one or both DM initial states are replaced by the other fermions $\psi_{2,3}$ and $\psi^{\pm}$.

\noindent
All the possible DM annihilation channels have been included in our numerical study, performed through the package micrOMEGAs~\cite{Belanger:2006is}, which allows also for a proper treatment of s-channel resonances as well as coannihilations, relevant in some regions of the parameter space. For a better insight, we report, nevertheless, below, the expressions of some phenomenologically interesting channels, by making use of the velocity expansion, $\langle \sigma v \rangle$, retaining only the leading order terms. We start with the $\bar{f} f$ channel:

\begin{align}
& \langle \sigma v \rangle_{ff}=\frac{1}{2\pi}\sum_f n_c^f \sqrt{1-\frac{m_f^2}{m_{\psi_1}^2}}\left[\frac{|\xi_A^f|^2 |y_{A\psi_1 \psi_1}|^2 m_f^2 m_{\psi_1}^2}{v^2 (4 m_{\psi_1}^2-m_A^2)^2}\right.\nonumber\\
&\left.+\frac{m_f^2}{m_Z^4}|g_{Z\psi_1 \psi_1}^A|^2 |g_{Zff}^A|^2-2 \frac{m_f m_{\psi_1}}{v \,m_Z^2}\mbox{Re}\left[(\xi_A^f)^{*}(y_{A\psi_1 \psi_1})^{*}g_{Z\psi_1 \psi_1}^A g_{Zff}^A\right]\right].
\end{align}

\noindent
As evident the $s$-wave term receives contributions only from s-channel exchange of the pseudoscalar Higgs $A$ and of the $Z$ boson with the latter being, however, helicity suppressed and, hence, relevant, for DM masses close to the mass of the top-quark.

\noindent
In the regime $m_{\psi_1} < m_{H,A,H^{\pm}}$ the other relevant annihilation channels are in the $WW$, $ZZ$ and $Zh$ final states. Their cross-sections are given by:

\begin{align}
& \langle \sigma v \rangle_{WW}=\frac{1}{4 \pi}\sqrt{1-\frac{m_W^2}{m_{\psi_1}^2}}\frac{1}{m_W^4 (m_W^2-m_{\psi_1}^2-m_{\psi^{\pm}}^2)^2}\left[(|g_{W\psi_1}^V|^2+|g_{W\psi_1}^A|^2)^2 (2 m_W^4 (m_{\psi_1}^2-m_W^2))\right.\\
& \left.+2 |g_{W\psi_1}^V|^2 |g_{W\psi_1}^A|^2 m_{\psi^{\pm}}^2 (4 m_{\psi_1}^4+3 m_W^2-4 m_{\psi_1}^2 m_W^2))\right] 
\end{align}

\begin{align}
\langle \sigma v \rangle_{ZZ}=\frac{1}{4 \pi}\sqrt{1-\frac{m_Z^2}{m_{\psi_1}^2}}\sum_{i=1,3}\frac{1}{(m_Z^2-m_{\psi_1}^2-m_{\psi_i}^2)^2} (|g_{Z\psi_1 \psi_i}^V|^2+|g_{Z\psi_1 \psi_i}^A|^2) (|g_{Z\psi_1 \psi_j}^V|^2+|g_{Z\psi_1 \psi_j}^A|^2) (m_{\psi_1}^2-m_Z^2)
\end{align}

\begin{align}
& \langle \sigma v \rangle_{Zh}=\frac{1}{\pi}\sqrt{1-\frac{(m_h+m_Z)^2}{4 m_{\psi_1}^2}}\sqrt{1-\frac{(m_h-m_Z)^2}{4 m_{\psi_1}^2}}\left[\frac{|\lambda_{hAZ}|^2 |y_{A\psi_1 \psi_1}|^2}{m_Z^2(4 m_{\psi_1}^2-m_A^2)^2}\left(m_{\psi_1}^4-\frac{1}{2}m_{\psi_1}^2 (m_h^2+m_Z^2)+\frac{1}{16}(m_h^2-m_Z^2)^2\right)\right.\nonumber\\
&\left. \frac{1}{256 m_{\psi_1}^2 m_Z^6}\lambda_{hZZ}^2 |g_{Z\psi_1 \psi_1}^A|^2 \left(m_h^4+(m_Z^2-4 m_{\psi_1}^2)^2-2 m_h^2 (m_Z^2-4 m_{\psi_1}^2)\right)\right.\nonumber\\
&\left. -\frac{1}{2}\mbox{Re}\left[\lambda_{hAZ}^{*} y_{A\psi_1 \psi_1}^{*} \lambda_{hZZ}g_{Z\psi_1 \psi_1}^A\right]\frac{1}{m_{\psi_1}m_Z^4 (4 m_\chi^2-m_A^2)}\left(m_{\psi_1}^4-\frac{1}{2}m_{\psi_1}^2 (m_h+m_Z)^2+(m_h^2-m_Z^2)^2\right)\right]
\end{align}

where the trilinear couplings used above are given by~\cite{Kanemura:2004mg}:
\begin{align}
& \lambda_{hZZ}=\frac{m_h^2}{2v}\sin(\alpha-\beta)\nonumber\\
& \lambda_{hAZ}=\frac{m_A^2-m_h^2}{v}\cos(\alpha-\beta)\nonumber\\
& \lambda_{HAZ}=\frac{m_A^2-m_H^2}{v}\sin(\alpha-\beta)\nonumber\\
& \lambda_{hAA}=-\frac{1}{4v\sin 2\beta}\left \{\left[\cos(\alpha-3\beta)+3\cos(\alpha+\beta)\right]m_h^2-4 \sin 2 \beta\sin(\alpha-\beta)m_A^2-4 \cos(\alpha+\beta)M^2 \right \}\nonumber\\
& \lambda_{HAA}=-\frac{1}{4v\sin 2\beta}\left \{\left[\sin(\alpha-3\beta)+3\sin(\alpha+\beta)\right]m_H^2+4 \sin 2 \beta\cos(\alpha-\beta)m_A^2-4 \sin(\alpha+\beta)M^2 \right \}
\end{align}

\noindent
The s-wave contributions to the $WW$ and $ZZ$ cross-sections are mostly determined by t-channel exchange of the new neutral and charged fermions, s-channel exchange of the Higgs states is present only in the velocity dependent term. The annihilation cross-section into $Zh$ receives an additional contribution, with respect to the ``minimal'' singlet-doublet model, from s-channel exchange of the pseudoscalar Higgs $A$.

\noindent
In this paper we will also consider the case that the mass of the DM is above the one of some new Higgs states. In particular we will consider the case of a light-pseudoscalar $A$. It is then useful to provide as well some estimates of the annihilation channels featuring $A$ and other light states, as the $h$ and the $Z$ boson, as final states:
\begin{align}
& \langle \sigma v \rangle_{ZA}=\frac{v_{\psi}^2}{16 \pi m_Z^2}\sqrt{1-\frac{(m_A-m_Z)^2}{4 m_{\psi_1}^2}}\sqrt{1-\frac{(m_A+m_Z)^2}{4 m_{\psi_1}^2}}\left(16 m_{\psi_1}^4-8 m_{\psi_1}^2 (m_Z^2+m_A^2)+(m_Z^2-m_A^2)^2\right)\nonumber\\
& \times {\left[\frac{\lambda_{hAZ}y_{h\psi_1 \psi_1}}{(4 m_{\psi_1}^2-m_h^2)}+\frac{\lambda_{HAZ}y_{H\psi_1 \psi_1}}{(4 m_{\psi_1}^2-m_H^2)}\right]}^2
\end{align}

\begin{align}
& \langle \sigma v \rangle_{hA}=\frac{1}{16\pi}\sqrt{1-\frac{(m_h+m_A)^2}{4 m_{\psi_1}^2}}\sqrt{1-\frac{(m_h-m_A)^2}{4 m_{\psi_1}^2}}\left[\frac{\lambda_{hAA}^2 y_{A\psi_1 \psi_1}^2}{(4 m_{\psi_1}^2-m_A^2)^2}+\frac{1}{4}\frac{\lambda_{hAZ}^2 g_{Z\psi_1 \psi_1}^2 (m_A^2-m_h^2)^2)}{(4 m_{\psi_1}^2-m_Z^2)^2}\right.\nonumber\\
& \left.\sum_{i,j=1,3} \frac{y_{A\psi_1 \psi_i}y_{A\psi_1 \psi_j}^{*}y_{h\psi_1 \psi_i}y_{h\psi_1 \psi_j}^{*}}{m_{\psi_1}^2 (m_A^2+m_h^2-2 m_{\psi_1}^2-m_{\psi_i}^2)^2 (m_A^2+m_h^2-2 m_{\psi_1}^2-m_{\psi_j}^2)^2 }\right.\nonumber\\
&\left. \times\left(m_A^4+m_h^4-8 m_{\psi_1}m_{\psi_j}m_h^2+16 m_{\psi_i}m_{\psi_j}m_{\psi_1}^2-2 m_A^2 (m_h^2-4 m_{\psi_1}m_{\psi_j}) \right)\right.\nonumber\\
&\left. \mbox{Re}\left[\lambda_{hAA}^{*}y_{A\psi_1 \psi_1}^{*}y_{h\psi_1 \psi_1}^{*}\lambda_{hAZ}g_{Z\psi_1 \psi_1}^A\right]\frac{(m_A^2-m_h^2)}{m_Z^2 m_{\psi_1}}\right.\nonumber\\
&\left.+\frac{2}{m_{\psi_1^2}}\mbox{Re}\left[\lambda_{hAA}^{*}y_{A\psi_1 \psi_1}^{*}y_{h\psi_1 \psi_1}^{*}y_{h\psi_1 \psi_i}y_{A\psi_1 \psi_i}\right]\frac{(m_A^2 m_{\psi_1}-m_h^2 m_{\psi_1}+4 m_{\psi_i}m_{\psi_1}^2)}{(m_A^2+m_h^2-2 m_{\psi_1}^2-2 m_{\psi_i}^2)(4 m_{\psi_i}^2-m_A^2)}\right.\nonumber\\
&\left. +\frac{1}{2}\sum_{i=1,3}\mbox{Re}\left[\lambda_{hAZ}^{*}g_{Z\psi_1 \psi_1}^{*}y_{h\psi_1 \psi_i}y_{A\psi_1 \psi_i}\right]\frac{(m_A^2-m_h^2)^2+4 m_{\psi_1}m_{\psi_i} (m_A^2-m_h^2)}{m_{\psi_1}^2 m_Z^2(m_A^2+m_h^2-2 m_{\psi_1}^2-2 m_{\psi_i}^2)}\right]
\end{align}

\begin{align}
& \langle \sigma v \rangle_{AA}=\frac{v_\psi^2}{128\pi}\sqrt{1-\frac{m_A^2}{m_{\psi}^2}}\left[{\left[\frac{\lambda_{AAh}y_{h\psi_1 \psi_1}}{(4 m_{\psi_1}^2-m_h^2)}+\frac{\lambda_{AAH}y_{H\psi_1 \psi_1}}{( m_{\psi_1}^2-m_H^2)}\right]}^2+\frac{16}{3}|y_{A\psi_1 \psi_1}|^2\frac{m_{\psi_1}^2 (m_{\psi_1}^2-m_A^2)^2}{(2 m_{\psi_1}^2-m_A^2)^4}\right.\nonumber\\
&\left. -\frac{8}{3}|y_{A\psi_1 \psi_1}|^2 \frac{m_{\psi_1}(m_{\psi_1}^2-m_A^2)}{(2 m_{\psi_1}^2-m_A^2)^2}\left[\frac{y_{h\psi_1 \psi_1}\lambda_{hAA}}{(4 m_{\psi_1}^2-m_h^2)}+\frac{y_{H\psi_1 \psi_1}\lambda_{HAA}}{(4 m_{\psi_1}^2-m_H^2)}\right]\right]
\end{align}

\noindent
Despite the velocity supression ($v_{\psi}^2$) the $AA$ and $ZA$ channels can provide not neglible contribution because of the sizable trilinear scalar couplings.

\begin{figure}
\begin{center}
\subfloat{\includegraphics[width=0.5\linewidth]{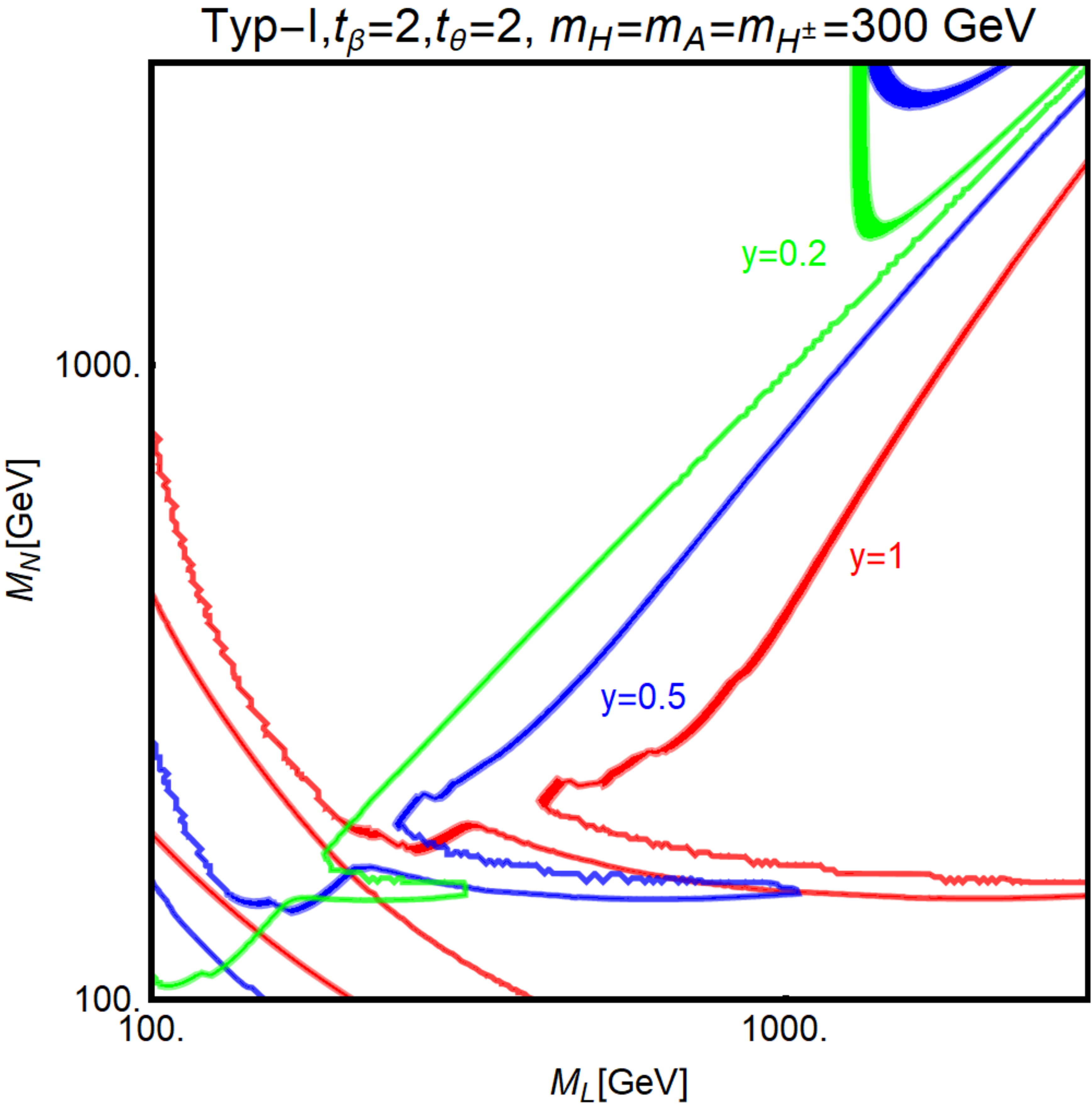}}
\subfloat{\includegraphics[width=0.5\linewidth]{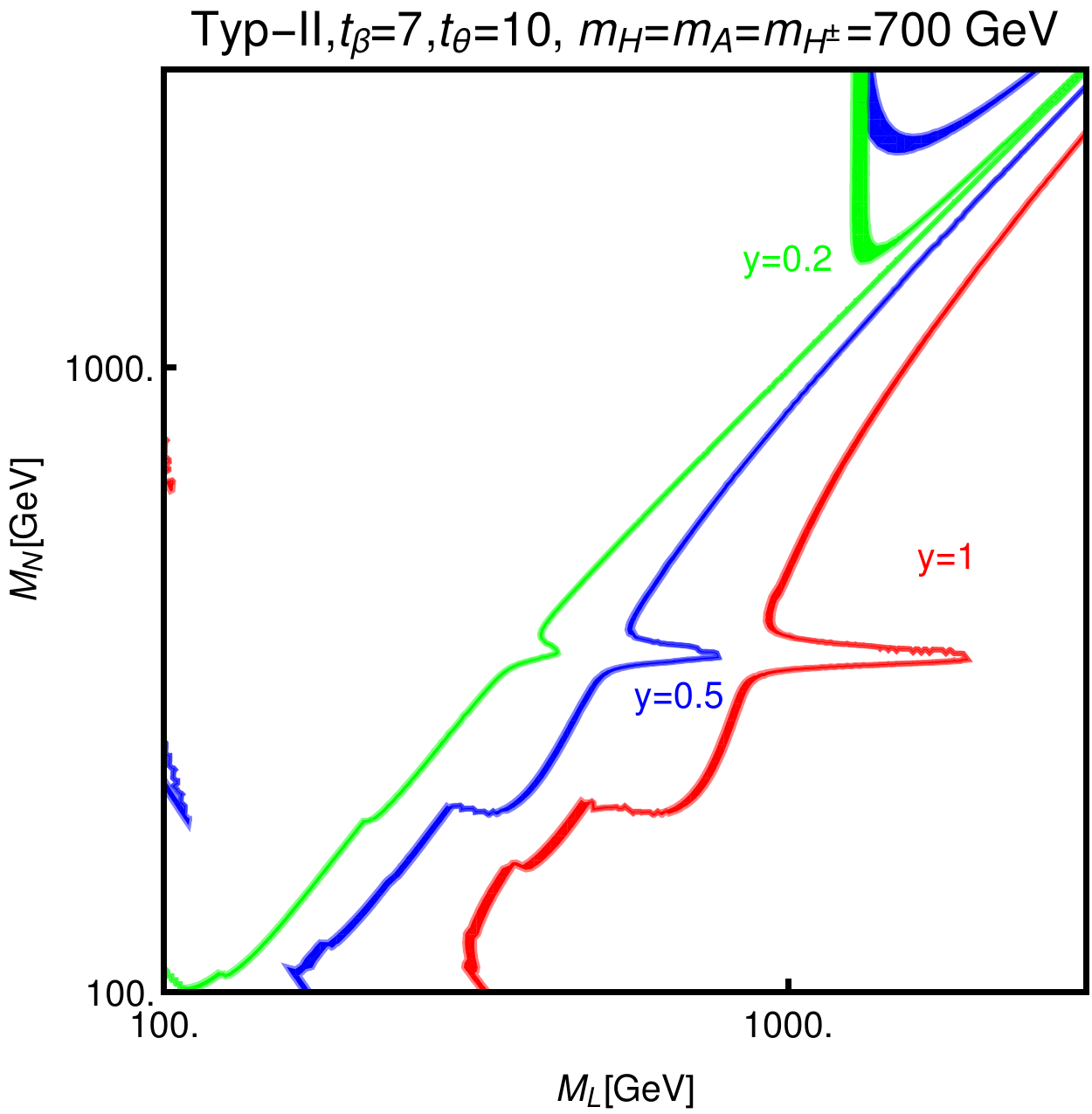}}
\end{center}
\caption{\footnotesize{Isocontours of the correct DM relic density in the bidimensional plane $(M_N,M_L)$ for type-I 2HDM (upper panels) and type-II 2HDM (lower panels), for three values of $y$, namely 0.2, 0.5 and 1, a some assignations of the other relevant parameters of the model, reported on top of the single panels.}}
\label{fig:relic}
\end{figure}

\noindent
We have reported in fig.~\ref{fig:relic} the isocontours corresponding to the correct DM relic density, $\Omega h^2 \approx 0.12$~\cite{Ade:2015xua}, in the bidimensional plane $(M_L,M_N)$ and for some assignations of the parameter of the model. For simplicity we have assumed $m_H=m_A=m_{H^{\pm}}$ and the alignment limit for the couplings of the Higgs bosons with SM states. In each panel of the figure we have considered three assignations of $y$, namely 0.2,0.5 and 1. The former assignation corresponds to the MSSM limit. As evidenced by the figure, for this assignation, the correct relic density is essentialy achieved in the ``well tempered'' regime $M_N \sim M_L$ until $M_L \sim 1.1 \mbox{TeV}$, where it is fully saturated by the annihilation into $WW$ pairs of a mostly doublet like DM, analogously to what happens for the MSSM higgsinos~\footnote{Notice that for DM masses above the TeV one should account for Sommerfeld enhanchment~\cite{Cirelli:2007xd,Cirelli:2009uv,Cirelli:2015bda}. As consequence of this the correct relic density is achieved for slightly higher masses, corresponding to $M_L \approx 1.4\,\mbox{GeV}$. Being the focus on this work on relic density a lower DM masses, we can neglect this effect.}. In this setup, the most relevant DM annihilation channels are controlled by the SM gauge couplings. A similar outcome would be then expected for lower values of $y$. For this reason we can focus, without loss of generality, to values $y \geq 0.2$. Notice that we have considered, in fig.~\ref{fig:relic}, higher values of the mass of the Higgses for the Type-II model. This is due to the stronger constraints, with respect to the type-I model, with from searches of new Higgs bosons at collider and in low energy phenomena.

\subsection{Direct Detection}

\noindent
In the scenario considered the DM features both spin independent and spin dependent interactions with nuclei. The former are induced, at three level, by t-channel exchange of the CP-even $h$ and $H$ states. The corresponding cross-section reads (for definiteness we report the case of scattering on protons):
\begin{equation}
\sigma_{\chi p}^{\rm SI}=\frac{\mu_\chi^2}{\pi}\frac{m_p^2}{v^2}|\sum_{q}f_q \left(\frac{g_{h\psi_1 \psi_1}\xi_h^q}{m_h^2}+\frac{g_{H\psi_1 \psi_1}\xi_H^q}{m_H^2}\right)|^2
\end{equation}
where $f_q^p$ are nucleon form factors (notice that the form factors corresponding to heavy quark are expressed in terms of the gluon form factor~\cite{Griest:1988yr,Drees:1993bu} as$f_c^p=f_c^p=f_t^p=\frac{2}{27}f_{TG}$)

\noindent
Spin dependent interaction are originated, instead, by interactions of the DM with the $Z$ boson, given the following cross-section:
\begin{equation}
\label{eq:SDcross}
\sigma_{\chi p}^{\rm SD}=\frac{3 \mu_{\chi p}^2}{m_Z^4}|g^A_{Z \psi_1 \psi_1}|^2\left[g_u^A \Delta_u^p+g_d^A \left(\Delta_d^p+\Delta_s^p\right)\right]^2
\end{equation}
where $\Delta_{u,d,s}^p$ are again suitable structure functions~\footnote{For the numerical values of all structure functions we have adopted the default assignations of the micrOMEGAs package.}.

\begin{figure}
\begin{center}
\subfloat{\includegraphics[width=0.5\linewidth]{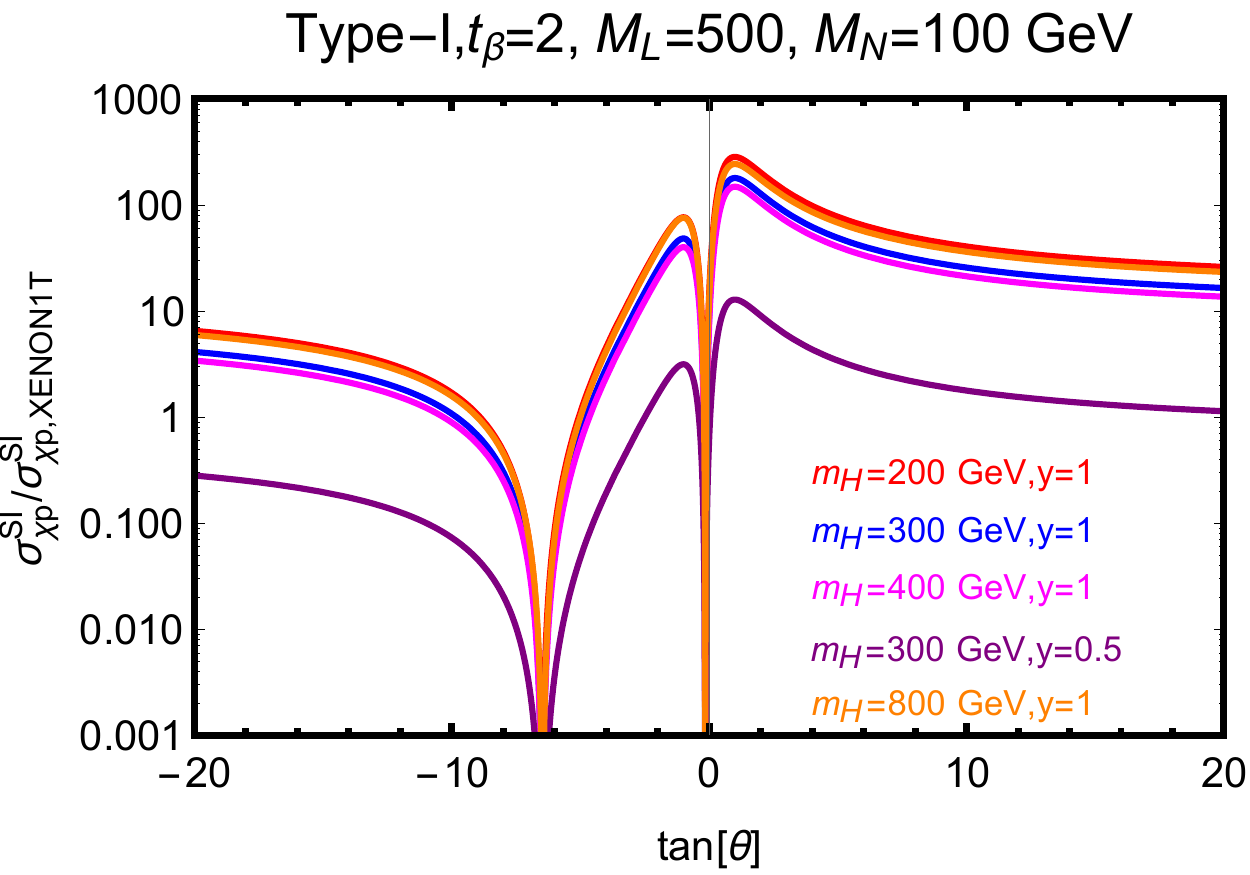}}
\subfloat{\includegraphics[width=0.5\linewidth]{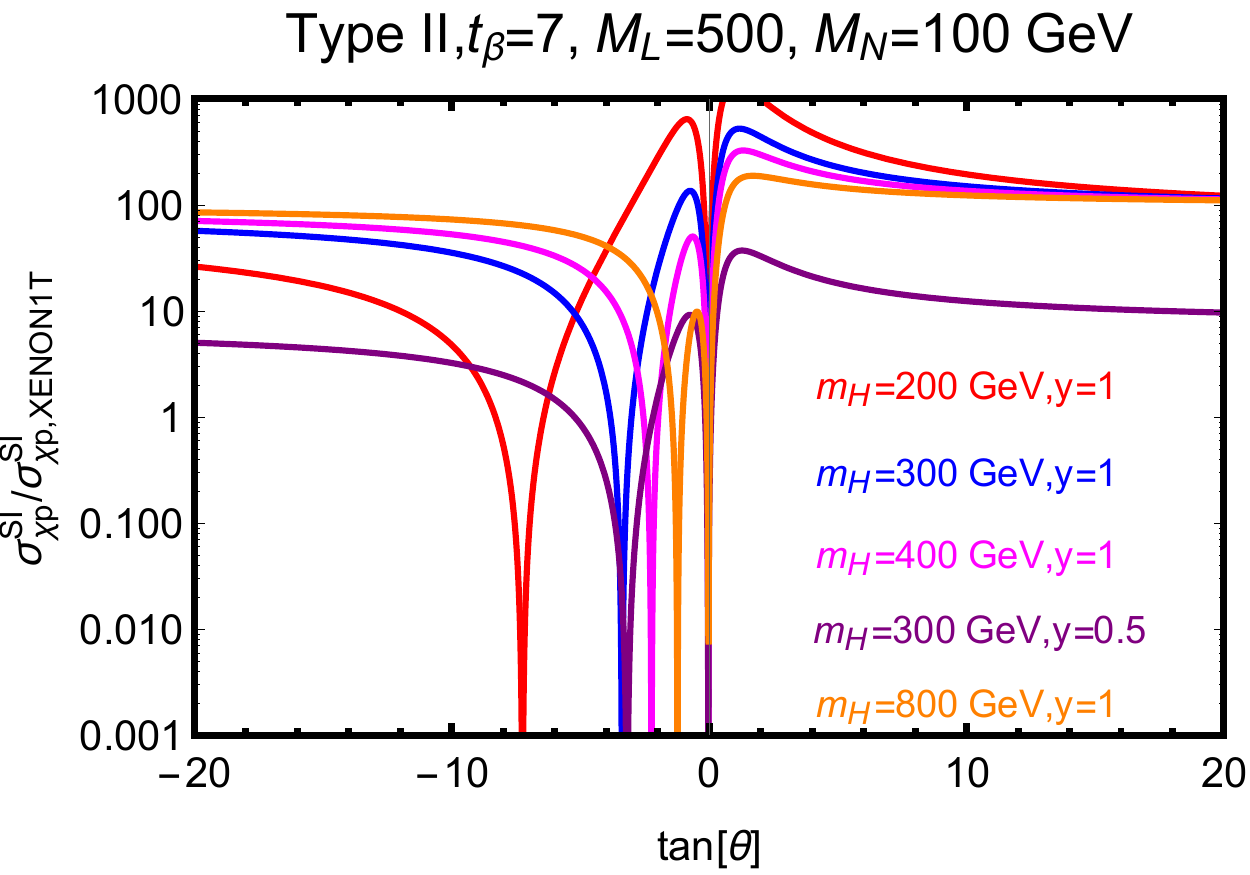}}
\end{center}
\caption{\footnotesize{Ratio of SI Direct Detection cross-section over the current limit value, as function of $\tan\theta$ for type-I (left panel), type-II (right panel), configurations and some assignations of $m_H$ and $y$, reported in the plots. In all cases we have fixed $M_L=500\,\mbox{GeV}$ and $M_N=100\,\mbox{GeV}$.}}
\label{fig:pBS}
\end{figure}

\noindent
As well known, Direct Detection constraints are mostly associated to spin independent interactions, because of the coherent enhanchment ($A^2$) occuring when these are evaluated at the nuclear, rather than nucleon, level. Hovever the different couplings of the DM with two mediators, the $h$ and $H$ can induce so called ``blind spots''~\cite{Cheung:2012qy,Huang:2014xua,Berlin:2015wwa,Choudhury:2017lxb}, due to a possible distructive interference between diagramms with $h$ and $H$ exchange. A blind spot in Direct Detection can be also created by a cancellation of the coupling of the DM with the $h$ state~\cite{Cheung:2013dua,Calibbi:2015nha} and by taking a moderately high, namely $\gtrsim 500\,\mbox{GeV}$, value of $m_H$. The occurrance of these blind spots, as function of $\tan\theta$ and some example assignations of the paramters, is shown in fig.~\ref{fig:pBS}. In regions where these strong cancellations occur, it might be necessary to take into account generally subdominant interactions like~\ref{eq:SDcross}. For this same reason we have included in our numerical study also one-loop corrections to SI cross-section arising from interaction of the DM with the $Z$ and $W$ bosons~\cite{Hisano:2010ct,Hisano:2010fy,Hisano:2011cs} and, eventually, from the pseudoscalar boson $A$~\cite{Ipek:2014gua,Arcadi:2017wqi,Sanderson:2018lmj} if this is light enough.

\subsection{Indirect Detection}

\noindent
As evidenced by the expressions provided in the previous subsection, some of the most relevant annihilation channels of the DM feature $s$-wave, i.e. velocity independent, annihilation cross-section. Thermal DM production can be thus tested, in our framework, also through Indirect Detection. Most prominent signals come from gamma-rays originating mainly from annihilations into $WW$, $ZZ$, $\bar t t$, $\bar b b$ and $\bar \tau \tau$. Particularly interesting would be, in this context the scenario of a light pseudo-scalar since it would be allow for a fit of the gamma-ray galactic center excess~\cite{Guo:2014gra,Cheung:2014lqa,Berlin:2015wwa}. DM interpretations of gamma-ray signals are, nevertheless, challenged by the exclusion limits from absence of evidences in Dwarf Spheroidal Galaxies (DSph)~\cite{Fermi-LAT:2016uux} as well as, since recently, searches in the Milky-Way Halo away from the GC~\cite{Chang:2018bpt} (these exclude, in particular, the $\bar b b$ interpratation of the GC excess).
In this work we will not attempt to provide a DM interpretation of the GC and focus, more conservatively, on the constraints on the DM annihilation cross-section.

\noindent
At the moment Indirect Detection constraints can probe thermal DM production up to DM masses of around 100 GeV and are rarely competitive with respect to Direct Detection constraints. In order to simplify the presentation of our results we will report Indirect Detection limits only when they are effectively complementary to other experimental searches while and omit them in the other cases.

\subsection{Invisible decays of the Higgs and of the Z}

\noindent
In the setup under considerations the DM is coupled, in pairs, both to the Higgs and to the Z boson. In the case it is lighter then $m_h/2,m_Z/2$, an invisible decay channel for the latter becomes accessible. This possibility is however experimentaly disfavored~\cite{Patrignani:2016xqp}. We have hence imposed in our analysis that, when the processes are kinematically allowed, the invisible branching ratio of the Higgs fullfills the upper bound $Br(h \rightarrow \mbox{inv})<0.2$ while, for what concerns the invisible width of the $Z$, $\Gamma(Z \rightarrow \mbox{inv}) < 2.3\,\mbox{MeV}$~\cite{Arcadi:2014lta}. 

\section{Results and discussion}

\begin{figure}
\begin{center}
\subfloat{\includegraphics[width=0.5\linewidth]{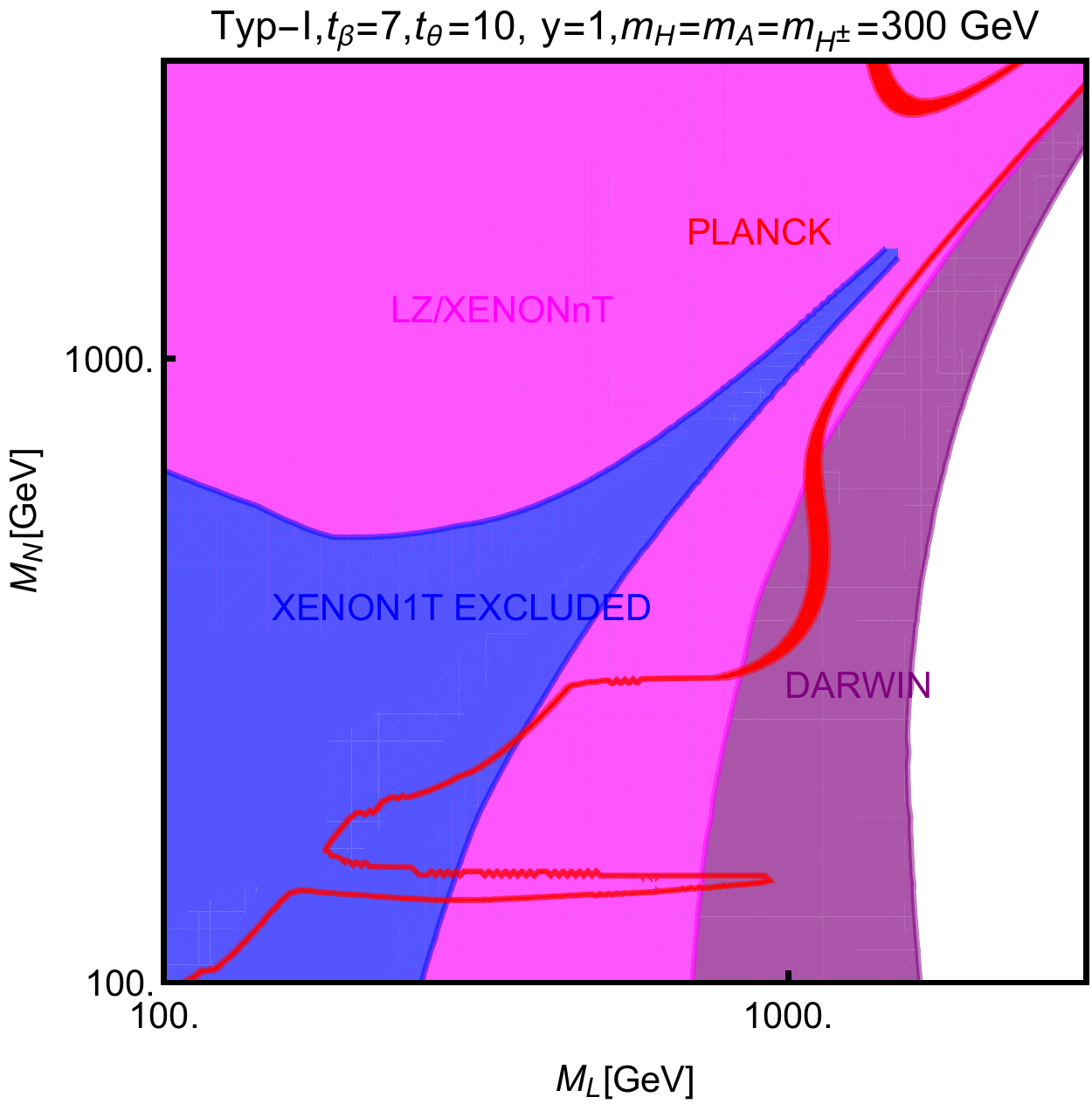}}
\subfloat{\includegraphics[width=0.5\linewidth]{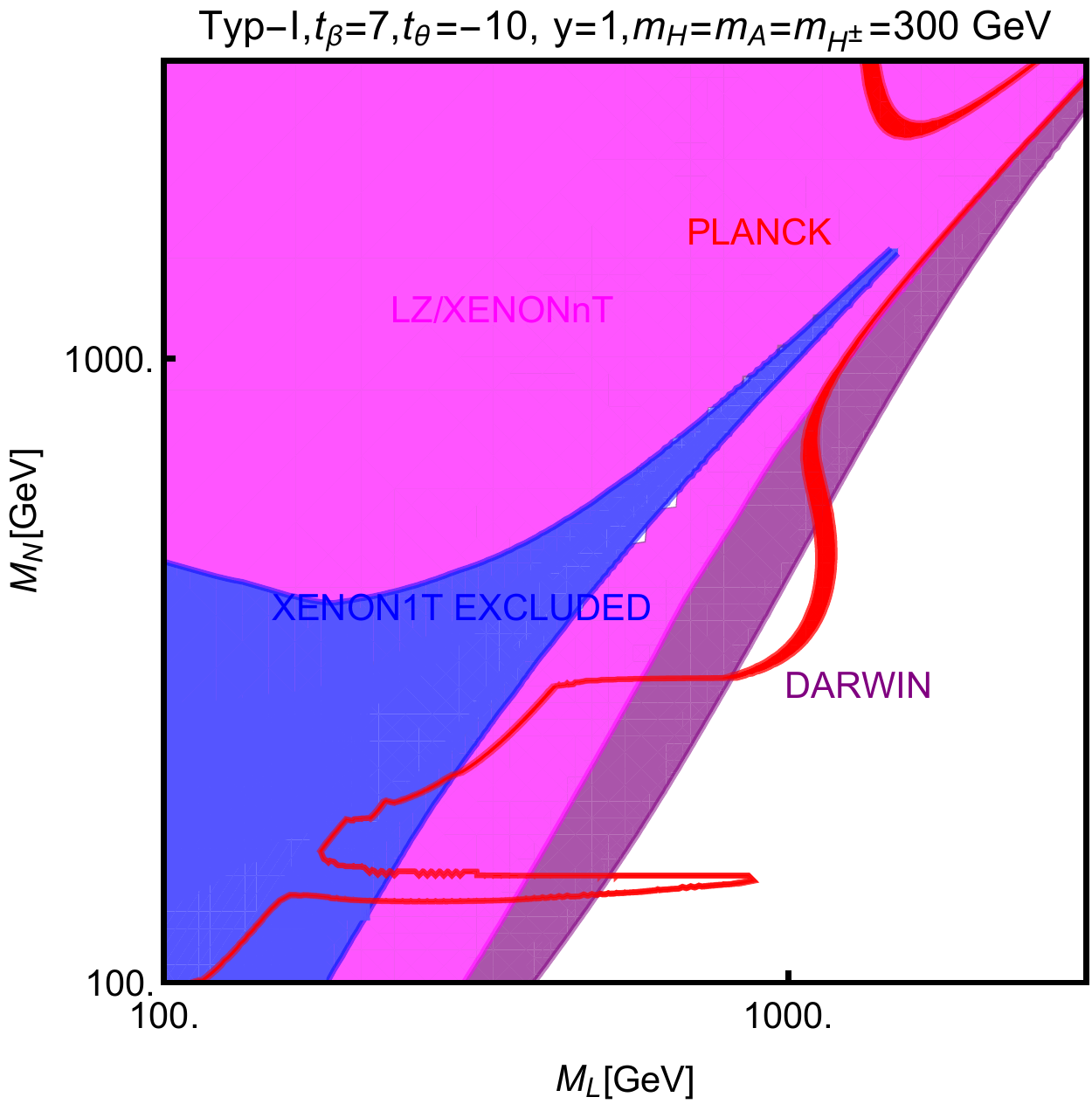}}\\
\subfloat{\includegraphics[width=0.5\linewidth]{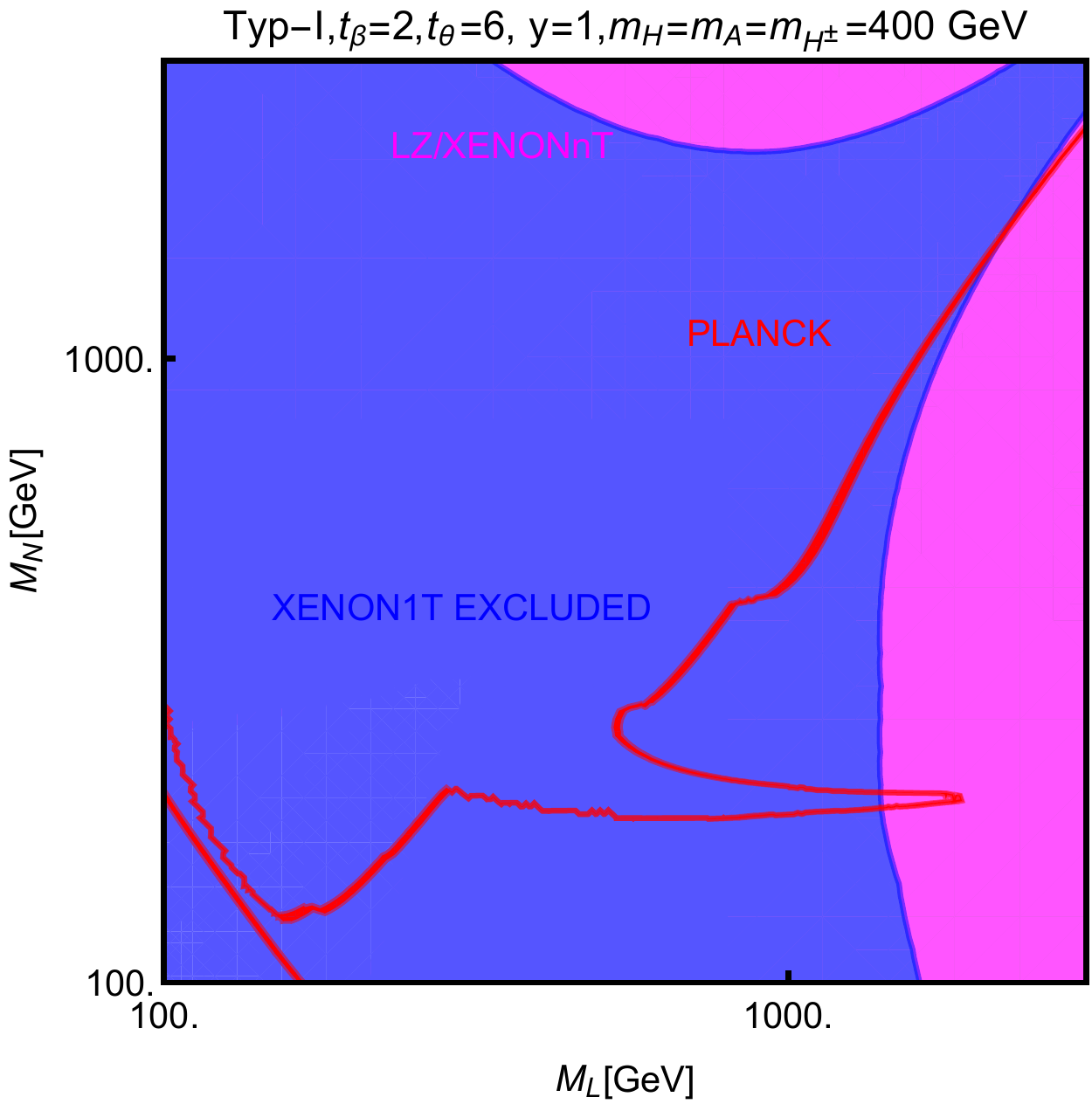}}
\subfloat{\includegraphics[width=0.5\linewidth]{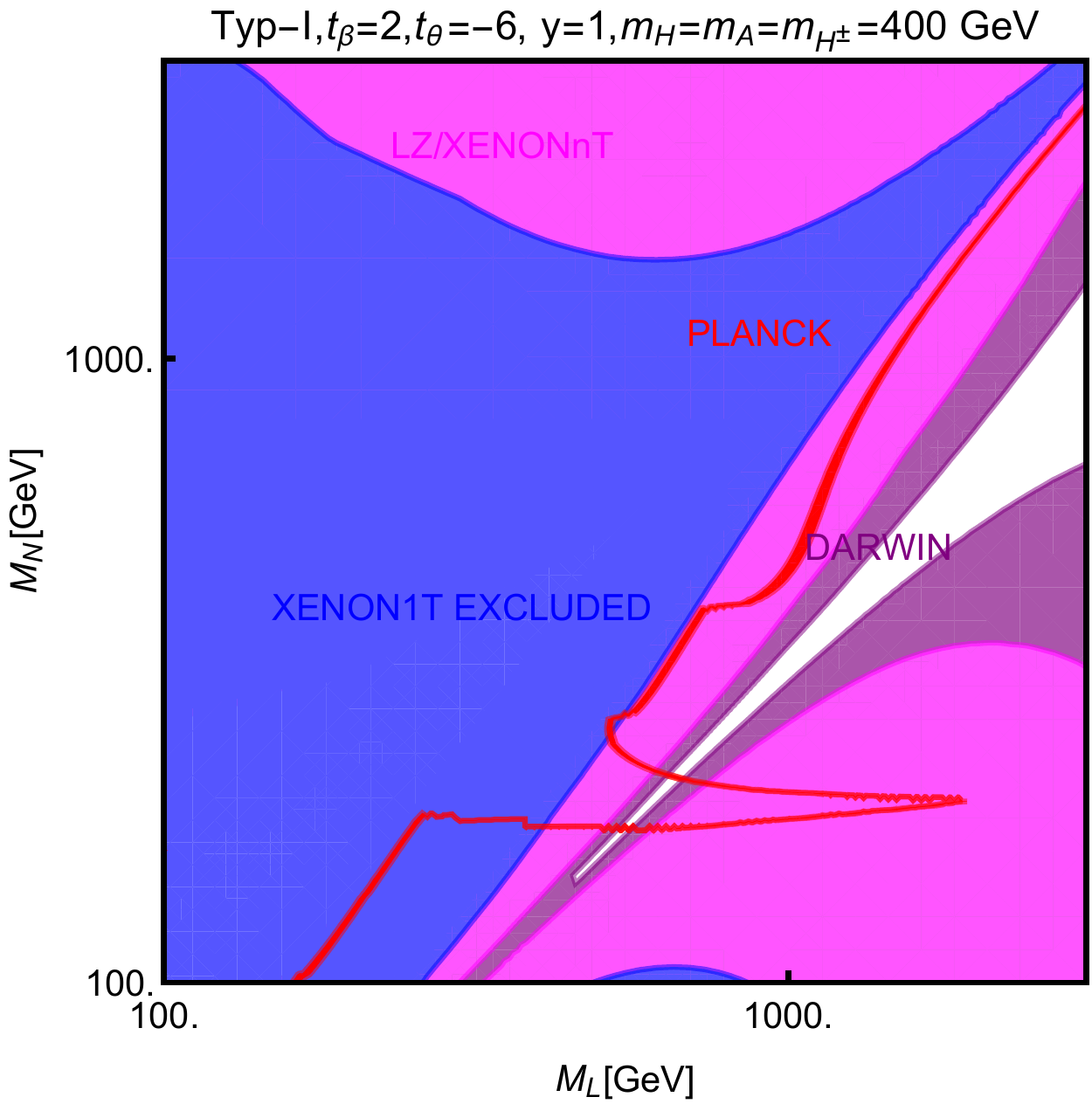}}
\end{center}
\caption{\footnotesize{Comparison between relic density constraints and limit/projected sensitivities, in the bidimensional plane $(M_L,M_N)$ in the type-I model for some assignations of $\theta,\tan\beta$ and $m_{H,A,H^{\pm}}$, reported on top of the panel. In each panel the red curve represents the isocontour of the correct relic density while the blue region is exclude by current constraints from XENON1T. The magenta and purple regions will be ruled out in case of null signals from, respectively, XENONnT/LZ and DARWIN.}}
\label{fig:fig2DtypI}
\end{figure}

\begin{figure}
\begin{center}
\subfloat{\includegraphics[width=0.5\linewidth]{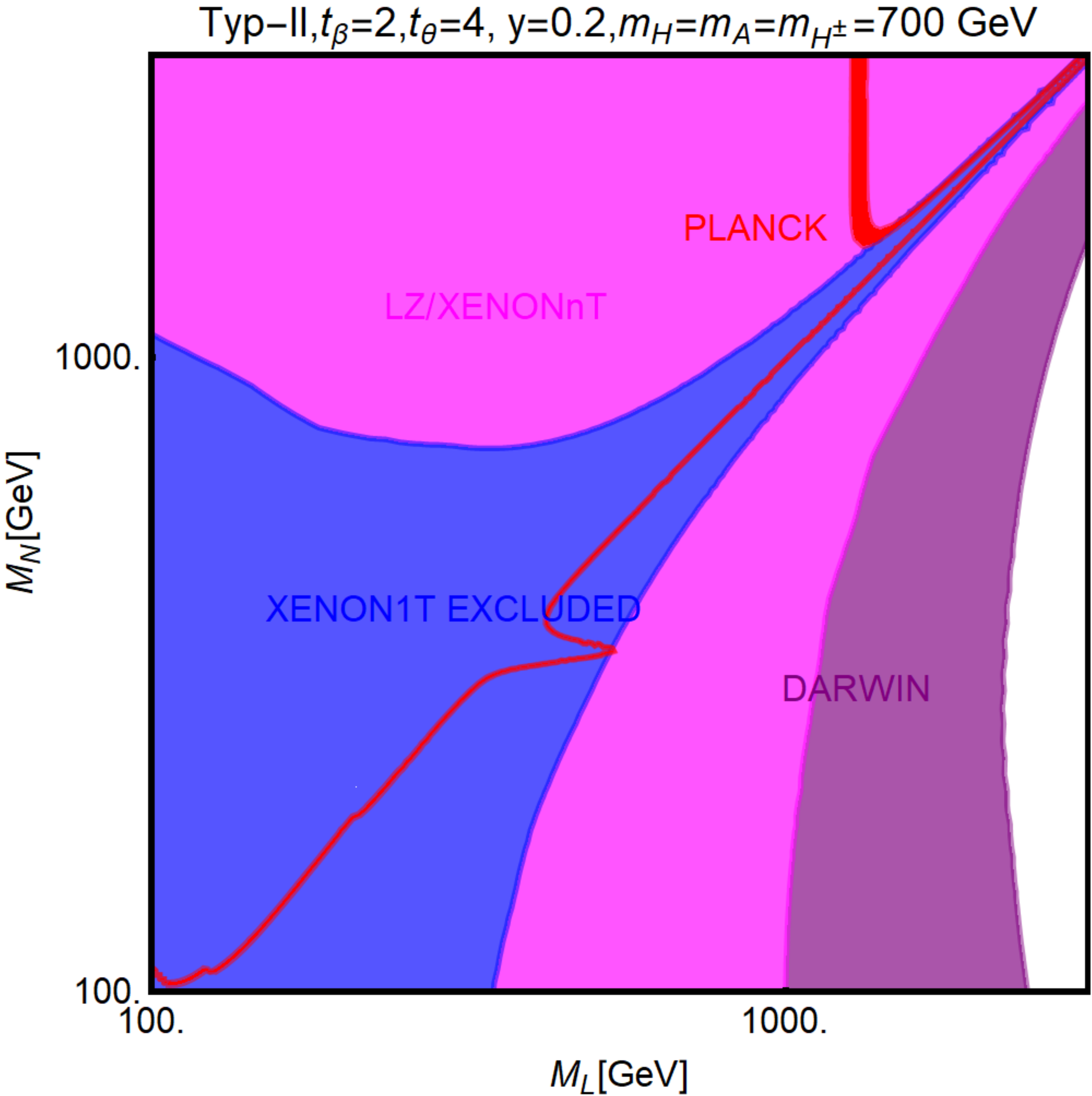}}
\subfloat{\includegraphics[width=0.5\linewidth]{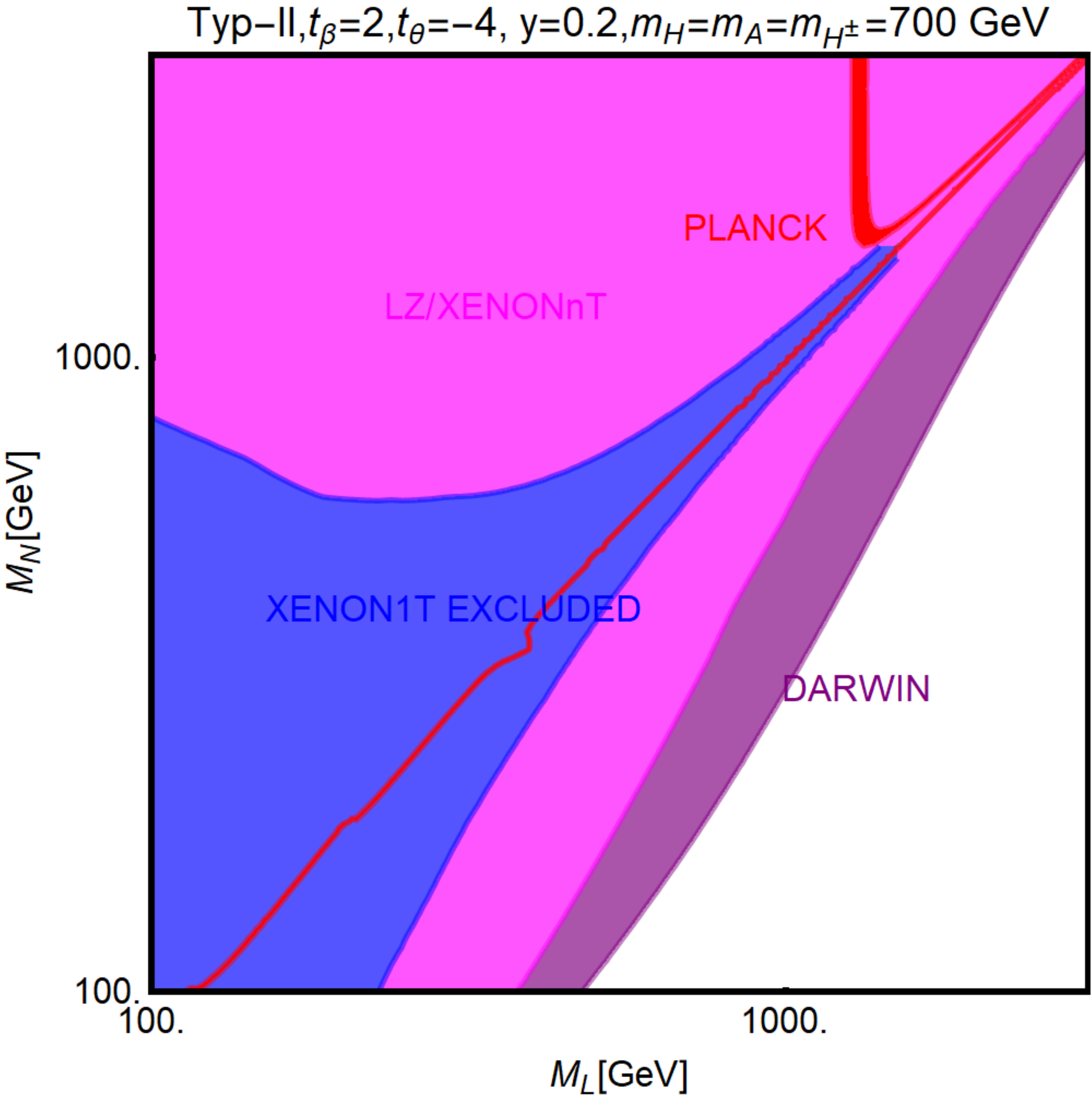}}\\
\subfloat{\includegraphics[width=0.5\linewidth]{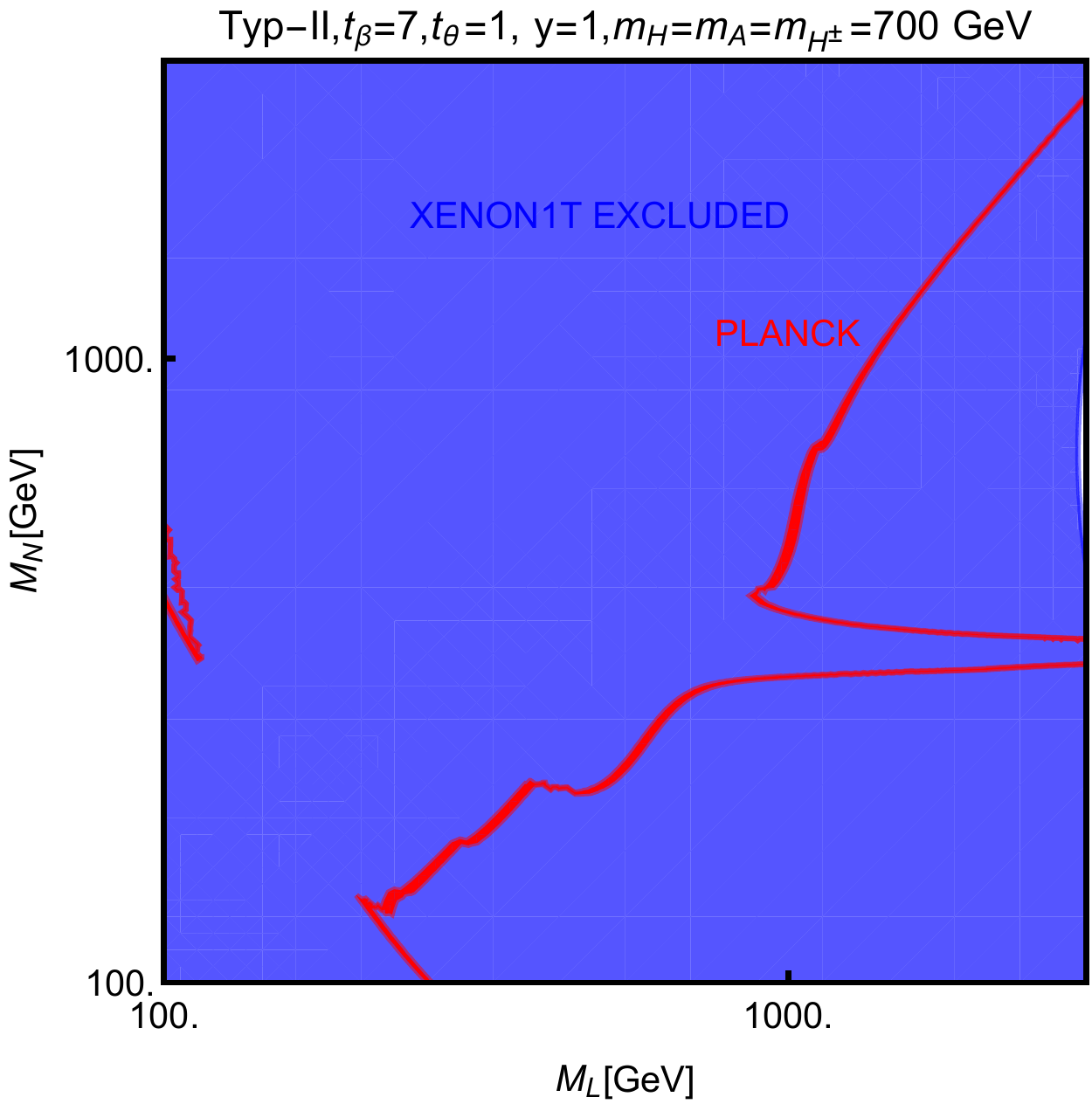}}
\subfloat{\includegraphics[width=0.5\linewidth]{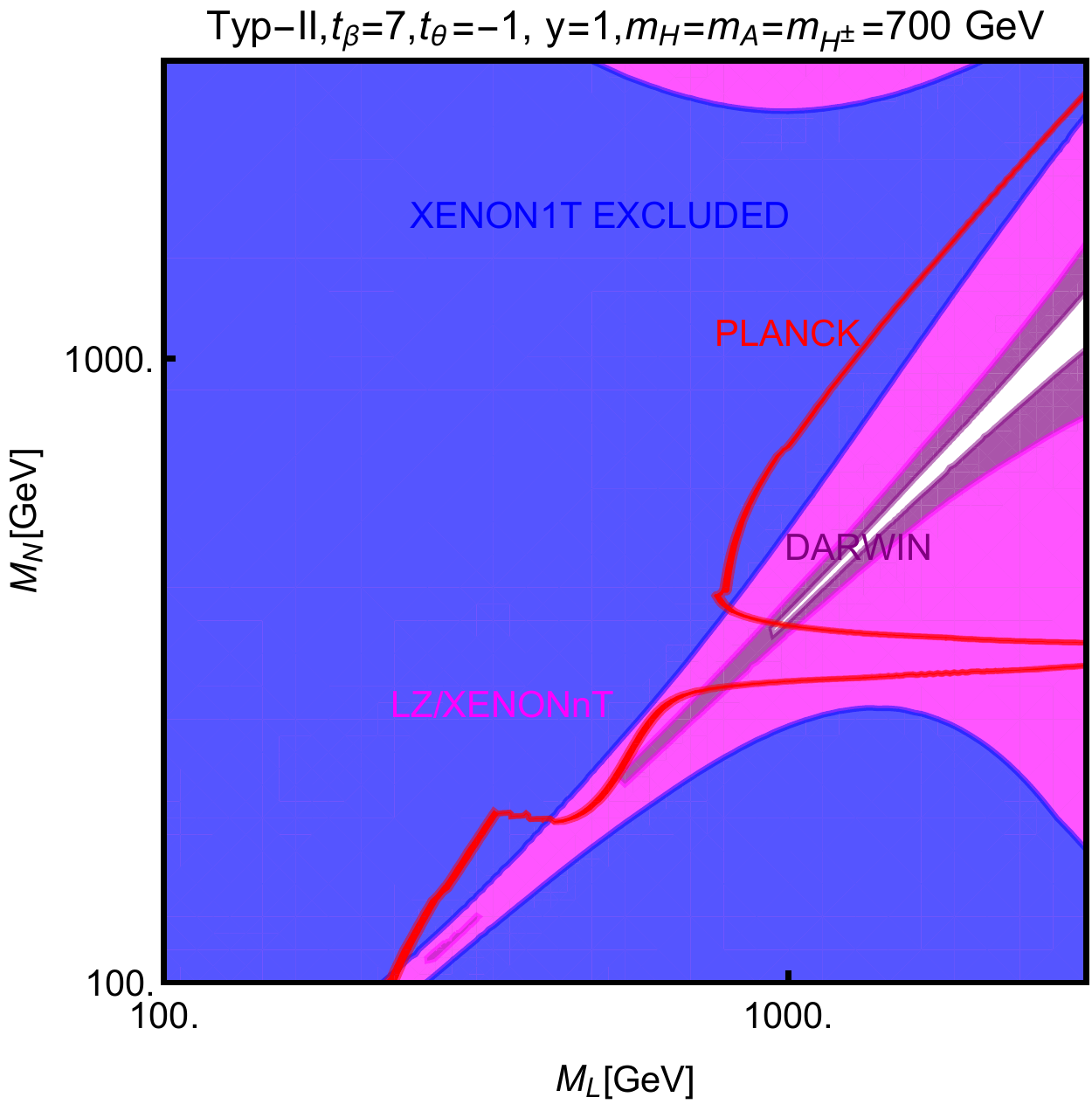}}
\end{center}
\caption{\footnotesize{The same as fig.~\ref{fig:fig2DtypI} but for type-II model and different assignations of the parameters of the model.}}
\label{fig:fig2DtypII}
\end{figure}

\noindent
We have now all the elements for examining in detail the constraints on the model under consideration. We show first of all in fig.~\ref{fig:fig2DtypI} and fig.~\ref{fig:fig2DtypII} the interplay between relic density and direct detection, in the bidimensional plane $(M_L,M_N)$, for, respectively, type I and type II scenarios. We have, again, focussed on some specific assignations of the other parameters of the theory and assumed, for simplicity, degenerate masses for the new bosons as well as the alignment limit. As we will see, constraints from Direct Detection are extremely strong, hence we focussed on $M_N,M_L > 100\,\mbox{GeV}$ in order to avoid the regions of maximal sensitivity for these experiments ($M_L <100\,\mbox{GeV}$ would be in any case forbidden by LEP limits on production of new charged particles). In each plot the parameter space corresponding to the correct relic density, represented by the red iso-contours, is compared with the excluded region (blue) by current limits from Direct Detection, essentially determined by XENON1T~\cite{Aprile:2017iyp}, as well as the projected sensitivities from XENONnT/LZ~\cite{Aprile:2015uzo,Szydagis:2016few} (magenta, given the similar sensitivity we are assuming the same projected excluded region for both experiments) and DARWIN~\cite{Aalbers:2016jon}(purple). As evident, the type-II model is extremely constrained, even once considering $y=0.2$ and relatively high masses of the new Higgs bosons. The only possibility to have viable DM is to rely on specific assignations of $\theta$, as shown in the right panels of fig.~\ref{fig:fig2DtypII}, corresponding to blind spots in the Direct Detection cross-section. These regions of the parameter space will be, however, completely probed by upcoming Direct Detection expermiments. The situation is better in the case of type-I model. As the first panel of fig.~\ref{fig:fig2DtypI} shows, it is possible to evade DD limits, even without relying on a blind spot configuration, at moderate values of $\tan\beta$, achieving a suitable suppression of the interactions of the DM with SM fermions, while still having a viable relic density, for $O(1)$ values of $y$, thanks to the annihilations into the Higgs states, which can be chosen still relatively light, thanks to the relatively weak bounds on the type-I 2HDM. Also in this case, however, negative signals from next generation Direct Detection experiments, would likely exclude thermal DM.  

\noindent
The type-I model offers another attractive possibility to evade direct detection constraints consisting into a light CP-odd boson $A$. In such a case, indeed, it is possible to achieve a sizable s-wave dominated annihilation cross-section of the DM into SM fermions, without strong additional contribution to the scattering cross-section since interactions mediated by a pseudoscalar are momentum suppressed (at least at the tree level). The DM annihilation cross-section can be also enhanced by the presence of $hA$, $AA$ and $ZA$ final states. Moreover the presence of a velocity independent cross-section would allow indirect detection as complementary probe and possibly to fit the GC excess~\cite{Cheung:2014lqa,Berlin:2015wwa}. Sizable constraints from negative gamma-ray signals from DSph would be present though.

\begin{figure}
\begin{center}
\subfloat{\includegraphics[width=0.5\linewidth]{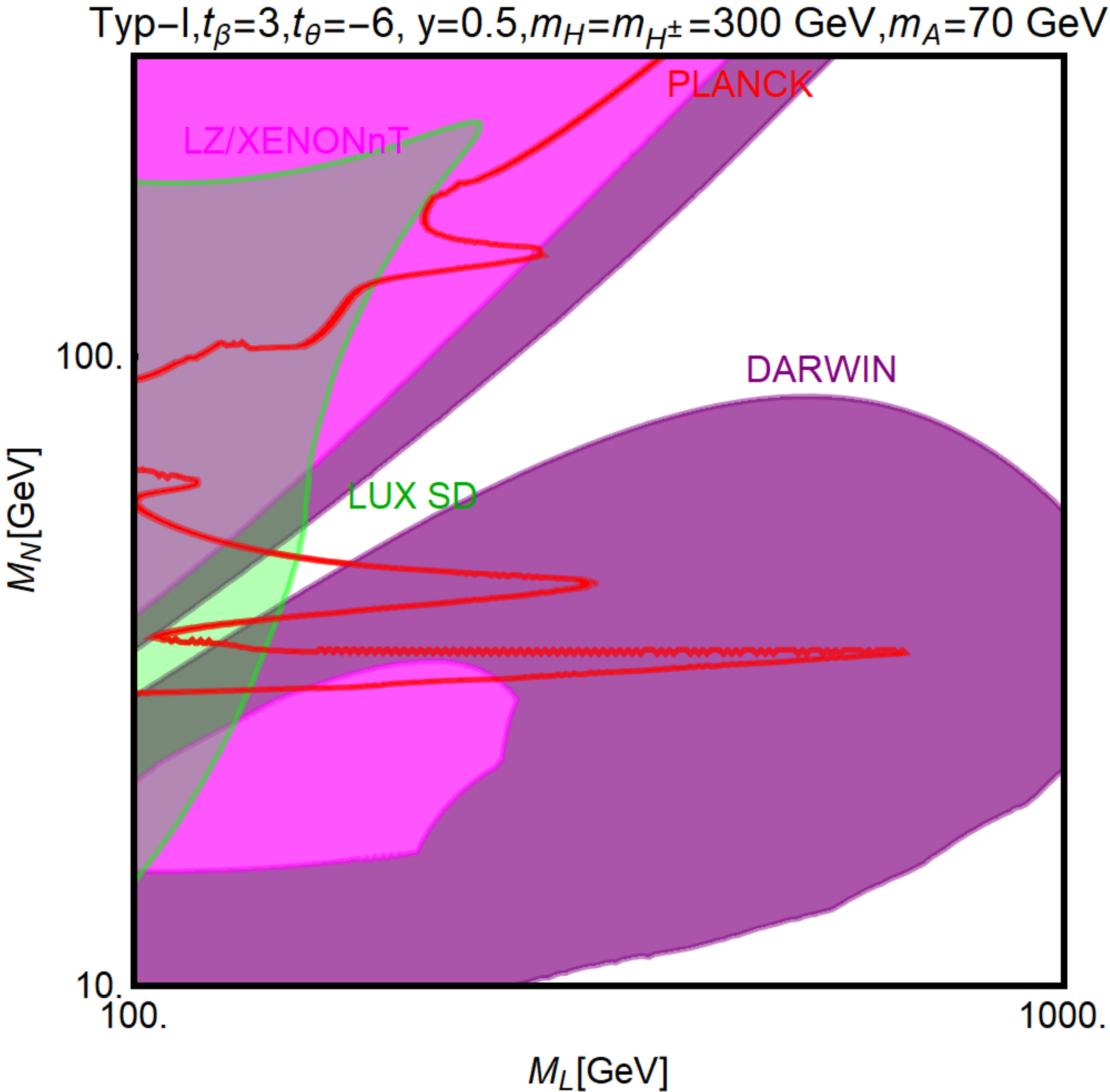}}
\subfloat{\includegraphics[width=0.5\linewidth]{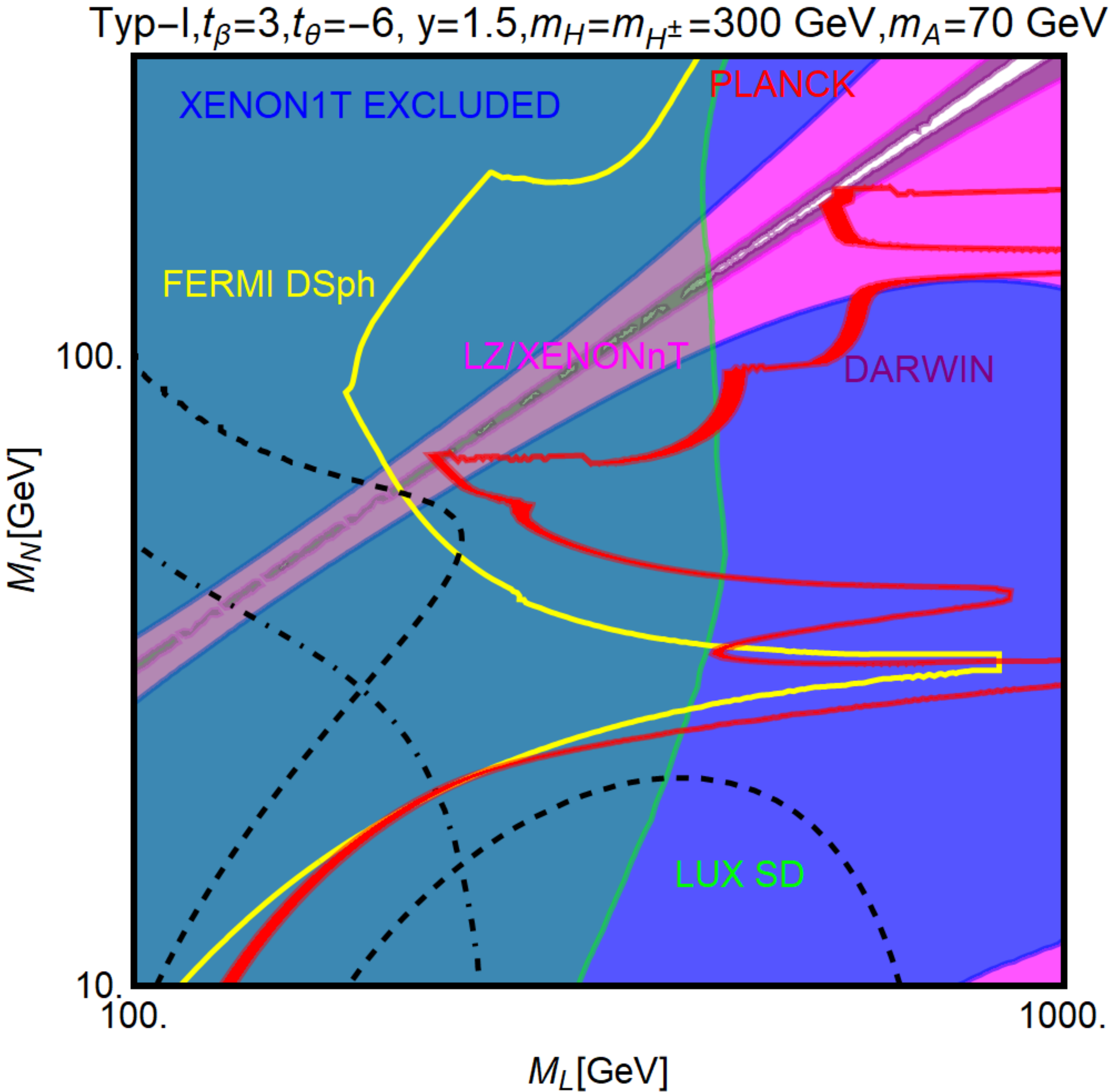}}\\
\subfloat{\includegraphics[width=0.5\linewidth]{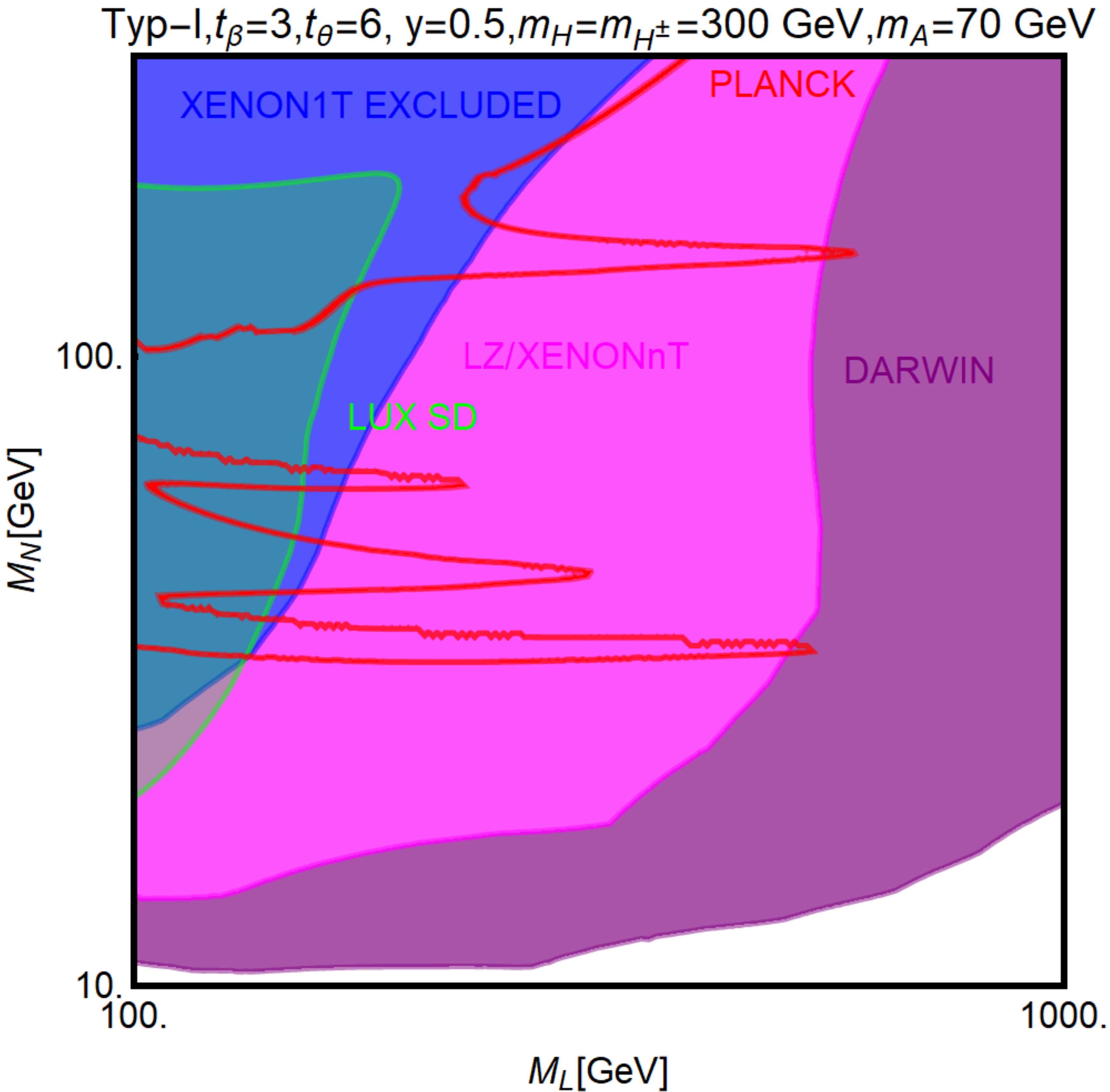}}
\subfloat{\includegraphics[width=0.5\linewidth]{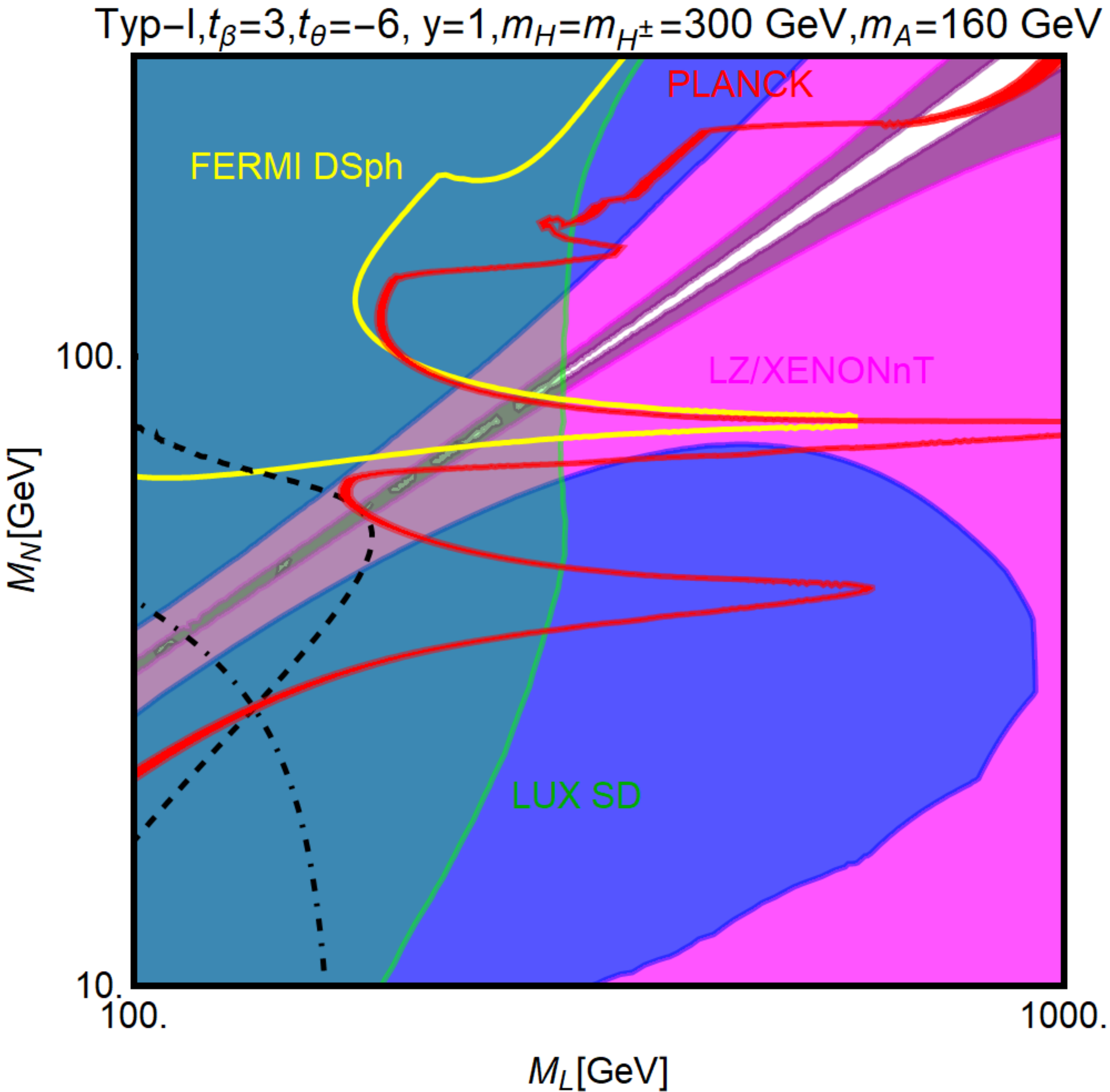}}
\end{center}
\caption{\footnotesize{Summary of DM constraints for some realizations of the type-I model with low mass pseudoscalar Higgs. As usual, the red contours correspond to the correct relic density while the blue/magenta/purple regions correspond to the present/future exclusions by XENON1T/XENONnT/DARWIN. In addition we have reported in green the excluded regions from SD limits, as provided by LUX. In the right panels the regions within the dashed (dot-dashed) black lines are excluded by the constraints on the invisible width of the Higgs ($Z$) boson while the regions inside the yellow contours are excluded by constraints from searches of $\gamma$-rays in DSph by FERMI.}}
\label{fig:lArelic}
\end{figure}

\noindent
We have then shown in fig.~\ref{fig:lArelic}, the combination of the DM constraits, for some parameter assignations, in the case of a light pseudoscalar Higgs $A$. As evident, the low mass mediator allows to achieve the correct relic density for DM masses below 100 GeV. At this low values, constraints from SI interactions are complemented by one from SD interactions (green regions in the plot) as well as from the invisible decay width of the $h$ and $Z$ bosons. Given the $t\beta$ suppression of the coupling of the extra Higgs bosons with SM fermions, Indirect Detection cannot efficiently probe the scenario under consideration unless values of $y$ above 1 (see second panel of fig.~\ref{fig:lArelic} are considered). 

\noindent
Regions of parameters space complying with all observational constraints are nevertheless present. These regions will be, however, fully probed and possibly ruled out by forthcoming Direct Detection experiments.

\section{Conclusions}

\noindent
We have performed an extensive analysis of the DM phenomenology of a model with singlet-doublet Dark Matter coupled with a two doublet Higgs sector. We have considered two scenarios for the couplings of SM and new fermions with the Higgs doublets resembling the conventional type-I and type-II 2HDM. In all cases the most competive constraints come from limits from Direct Detection. In the case of the type-II model these can be evaded only by invoking parameter assignations inducing blind-spots in the couplings responsible for Direct Detection. In the case of type-I model is instead possible to evade Direct Detection constraints even without relying on blind spots. The type-I model presents the additional interesting possibility of a light pseudoscalar Higgs boson.  

\noindent
For all the considered scenarios, next future direct detection facilities will full probe the viable region for thermal DM relic density.

\section*{Acknowledgements}

\noindent
We thank Federico Mescia and Olcyr Sumensari for the frutiful discussions. We are also indebted with Luca di Luzio for the valuable comments on the draft.

\bibliography{bibfile}{}

\end{document}